\documentclass[lettersize,journal]{IEEEtran}
\usepackage{pifont}
\newcommand{\cmark}{\ding{51}}
\newcommand{\xmark}{\ding{55}}%
\usepackage{amsmath,amsfonts}
\usepackage{xcolor}
\usepackage{makecell}
\usepackage{subcaption}
\usepackage{dsfont}
\usepackage{algorithmic}
\usepackage{algorithm}
\usepackage{array}
\usepackage[caption=false,font=normalsize,labelfont=sf,textfont=sf]{subfig}
\usepackage{textcomp}
\usepackage{stfloats}
\usepackage{url}
\usepackage{verbatim}
\usepackage{graphicx}
\usepackage{cite}
\usepackage{bibunits}
\begin{document}
\begin{bibunit}[IEEEtran]

\title{Multi-Domain EEG Representation Learning with Orthogonal Mapping and Attention-based Fusion for Cognitive Load Classification}


\author{Prithila Angkan, Amin Jalali, Paul Hungler, Ali Etemad
\thanks{Prithila Angkan, Amin Jalali and Ali Etemad are with the Department of Electrical and Computer Engineering and Ingenuity Labs Research Institute, Queen’s University, Kingston, Ontario, Canada.
Emails: prithila.angkan, amin.jalali, ali.etemad@queensu.ca
Paul Hungler is with Ingenuity Labs Research Institute,
Queen’s University, Kingston, Ontario, Canada.
Email: paul.hungler@queensu.ca}
}

\markboth{}%
{Shell \MakeLowercase{\textit{et al.}}: A Sample Article Using IEEEtran.cls for IEEE Journals}

\maketitle

\begin{abstract}
We propose a new representation learning solution for the classification of cognitive load based on Electroencephalogram (EEG). Our method integrates both time and frequency domains by first passing the raw EEG signals through the convolutional encoder to obtain the time domain representations.   Next, we measure the Power Spectral Density (PSD) for all five EEG frequency bands and generate the channel power values as 2D images referred to as multi-spectral topography maps. These multi-spectral topography maps are then fed to a separate encoder to obtain the representations in frequency domain. Our solution employs a multi-domain attention module that maps these domain-specific embeddings onto a shared embedding space to emphasize more on important inter-domain relationships to enhance the representations for cognitive load classification. Additionally, we incorporate an orthogonal projection constraint during the training of our method to effectively increase the inter-class distances while improving intra-class clustering. This enhancement allows efficient discrimination between different cognitive states and aids in better grouping of similar states within the feature space. We validate the effectiveness of our model through extensive experiments on two public EEG datasets, CL-Drive and CLARE for cognitive load classification. Our results demonstrate the superiority of our multi-domain approach over the traditional single-domain techniques. Moreover, we conduct ablation and sensitivity analyses to assess the impact of various components of our method. Finally, robustness experiments on different amounts of added noise demonstrate the stability of our method compared to other state-of-the-art solutions.
\end{abstract}


\section{Introduction}
\IEEEPARstart{E}{lectroencephalography (EEG)} serves as a non-invasive method for measuring the electrical activities of the brain by placing electrodes on the scalp and forehead \cite{binnie1994electroencephalography}. 
Numerous studies have highlighted various factors influencing brain activity \cite{dai2022semi}, including cognitive load and affect \cite{ko2020vignet, antonenko2010using}. 
As a result, EEG signals can be recorded and leveraged in conjunction with machine learning and deep learning techniques for detecting and quantifying cognitive load \cite{pulver2023eeg} and emotions \cite{suhaimi2020eeg}. 
\textcolor{black}{Cognitive load is defined as the mental workload required to perform a task \cite{sweller2016cognitive}. 
It reflects the amount of cognitive resources consumed during information processing, decision-making, or problem-solving activities.
For instance, in the context of driving, any alteration in cognitive load can impact driving performance. 
Notably, heightened cognitive load tends to reduce performance, particularly among novice drivers, increasing the likelihood of accidents \cite{ross2014investigating}.}

\begin{figure}
  \includegraphics[width=1\columnwidth]{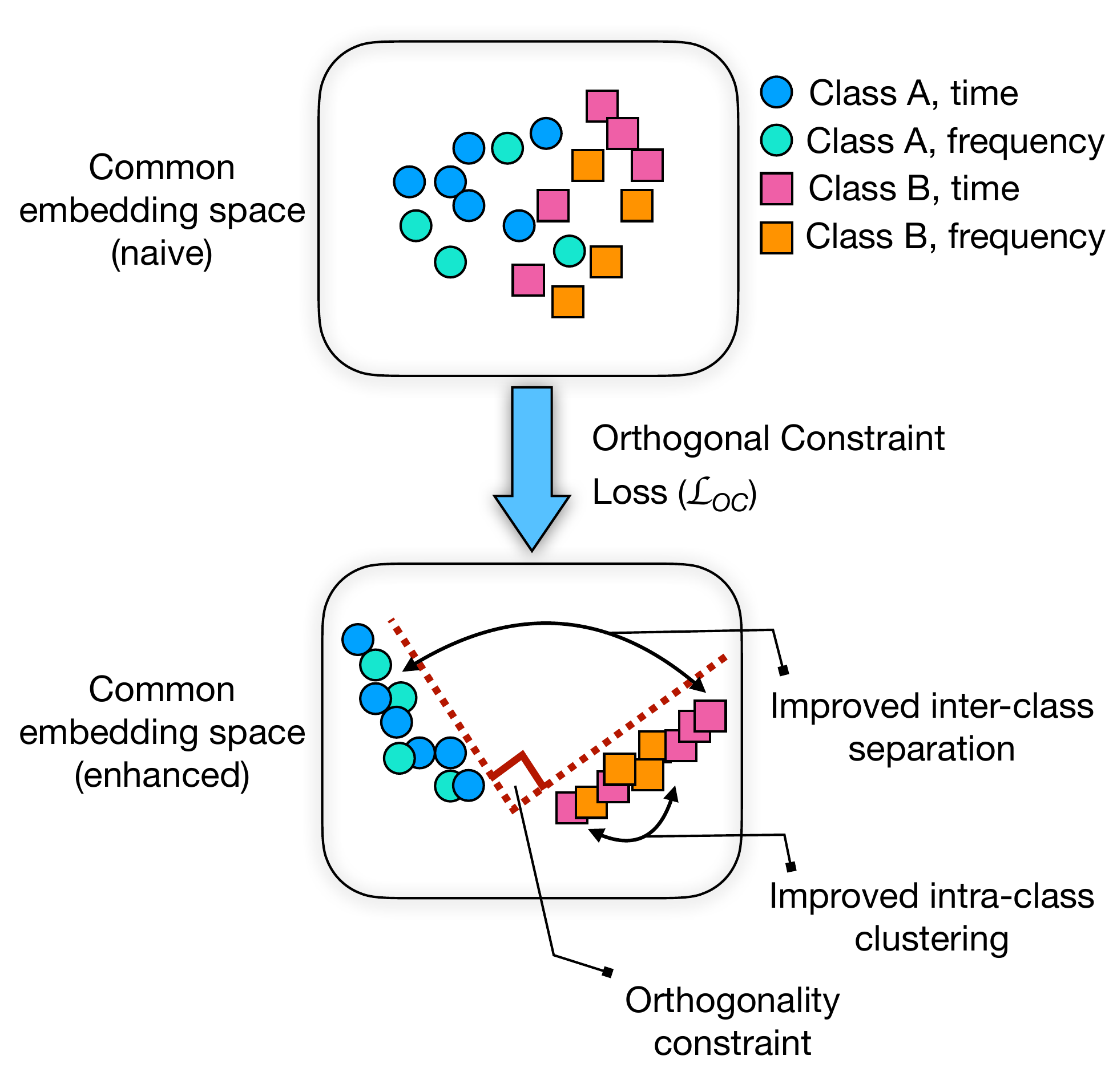}
  \caption{Overview of the orthogonality constraint used in our method, which improves inter-class separation and intra-class clustering for the fused embedding space.}
  \label{fig:oc_loss}
\end{figure}

Cognitive load can be assessed through a variety of modalities, including physiological signals \cite{chen2013automatic}, speech \cite{yin2008speech}, eye-tracking \cite{zagermann2016measuring}, and computer vision \cite{rahman2021vision}. Among these, EEG analysis has gained significant attention for its effectiveness in determining cognitive load \cite{angkan2024multimodal, antonenko2010using}. \textcolor{black}{More specifically, prior works demonstrate that cognitive load impacts the prefrontal cortex, frontal and parietal regions of the brain \cite {miller2000prefontral, gevins2000neurophysiological}. This include electrodes Fp1, Fp2, Fz, F3, F4, FC1, FC2, AF3, AF4, AF7, AF8, TP9, TP10, P3, P4, Pz. The Muse S headband has 4 electrodes AF7, AF8, TP9 and TP10, which are close to the electrodes of frontal and prefrontal cortex allowing the device to capture meaningful information. EEG signals, known for their high temporal resolution, can detect subtle changes in brain activity \cite{burle2015spatial}.} This makes them particularly well-suited for assessing alterations in cognitive load.

\textcolor{black}{Prior research suggested that cognitive load and emotion are closely linked, influencing each other both physiologically and computationally \cite{plass2019four,hawthorne2019well}. Physiological signals like EEG, ECG, and GSR reflect both cognitive and affective states and are widely used in deep learning-based estimation and many EEG analysis techniques (including multi-domain feature fusion and attention mechanisms)\cite{pulver2023eeg, bhatti2024attx}. Reviewing both domains is therefore essential to motivate our approach, which aims to address limitations and improve classification.}

\textcolor{black}{Prior works on EEG representation learning have either utilized features in time domain \cite{wan2020hybrideegnet} or frequency domain  \cite{peng2023ta, bai2023sect}. Despite the success of single-domain approaches in EEG representation learning \cite{pulver2023eeg, ng2019psd, al2017classification}, the integration of both time and frequency domains remains largely unexplored, especially in the context of cognitive load classification. We hypothesize that leveraging the complementary information from both domains has the potential to enhance the richness of the learned representations and improve overall system performance. Embeddings derived from different domains may possess different statistical properties in the representational space \cite{manzoor2023multimodality,liang2024quantifying}. Therefore, a primary challenge in devising an effective multi-domain approach for representation learning lies in the creation of a \textit{unified} yet distinctive embedding space \cite{hubert2017learning,pandey2022cross}.}

To address this gap, we propose a multi-domain representation learning framework for cognitive load that integrates EEG in both time and frequency domains within a dual-stream network. 
\textcolor{black}{Unlike prior EEG fusion methods that simply concatenate features from different domains, our approach enforces an \emph{orthogonal mapping} between the time-domain and frequency-domain feature spaces. Combining time-domain and frequency-domain EEG features yields a richer representation that improves classification performance, as demonstrated by our experimental results. The orthogonal mapping ensures that these features are complementary and non-redundant, enhancing the model’s generalization ability. The attention-based fusion further allows the model to adaptively emphasize the most relevant features in each domain, leading to robust cognitive load detection which is not explored in existing methods \cite{wan2020hybrideegnet,peng2023ta, bai2023sect}.}
This approach aims to generate enriched, fused embeddings by utilizing the unique and complementary characteristics of each domain. Our framework uses raw EEG time-series as the time domain representation, while Topological Mapping of EEG-based Brain Activity, referred to as multi-spectral topography maps, are used to represent the EEG information in frequency domain. Subsequent to encoding the data in each domain, we map the embeddings into a shared feature space, aligning them according to the labels via an attention module. The attention module \cite{chen2020multi,tu2023dctm} effectively captures the inter-domain relationships by emphasizing shared informative features from each domain, thereby enhancing the unified feature representation. Furthermore, we impose orthogonality constraints \cite{jalali2019atrial,ranasinghe2021orthogonal} to ensure that fused embeddings of different classes are orthogonal, thus maximizing inter-class separability, while embeddings from the same class remain similar, thereby minimizing intra-class separation (see Figure \ref{fig:oc_loss}). We evaluate the performance of our method on two public EEG-based cognitive load datasets, CLARE \cite{bhatti2024clare} and CL-Drive \cite{angkan2024multimodal}, where our method achieves state-of-the-art performances. Additionally, detailed ablation studies demonstrate the impact of each component of our solution.

The contributions of this paper are summarized as follows.
\begin{itemize}
\item To the best of our knowledge, this is the first study to integrate raw EEG time-series with multi-spectral topography maps within a multi-domain framework for cognitive load classification. This multi-domain strategy yields superior performance compared to utilizing a single domain of data.

\item We enhance the capabilities of our model by integrating a multi-domain attention mechanism to capture the complementary inter-domain relationships by concentrating on pertinent information from each domain. This generates an enhanced feature representation for processing EEG signals.

\item Additionally, we propose the incorporation of an orthogonal mapping constraint within the multi-domain structure. This ensures that fused embeddings of different classes are well-separated, thus maximizing inter-class distance. Concurrently, it maintains the similarity of embeddings within the same class, thereby minimizing intra-class distance. 
\end{itemize}

\section{Related Work} \label{Related Work}
In this section, we first provide an overview of EEG signal analysis for the classification of cognitive load. Prior works have also suggested that cognitive load is closely related to affect \cite{plass2019four, um2012emotional}. As a result, we also provide a brief overview of prior works on EEG-based affective computing.

\subsection{Cognitive load classification from EEG}
Utilizing EEG data to classify cognitive load levels in various tasks has gained attention in recent years \cite{yoo2023prediction, farkish2023evaluating, pulver2023eeg}. A thorough review of the literature reveals that a key inhibiting factor in the area of cognitive load classification with biological signals, especially EEG, is lack of various publicly available datasets. To our knowledge, CLARE \cite{bhatti2024clare} and CL-Drive \cite{angkan2024multimodal} are the only public datasets containing measured cognitive load and EEG signals along with other modalities. Among the works reviewed below, \cite{zarjam2013estimating, zarjam2015beyond, liu2023fusion, wang2023characterisation} have all collected new data to be used in the respective experiments, while \cite{pulver2023eeg} has used the CL-Drive dataset presented in \cite{angkan2024multimodal}.

In \cite{zarjam2013estimating}, seven levels of cognitive load were induced in 12 male participants using arithmetic tasks. The experiment was conducted in one session and the length of the session was 15 minutes. Wavelet coefficient entropy was then extracted for predicting cognitive load using an Artificial Neural Network (ANN) classifier.
Arithmetic task was also used in \cite{zarjam2015beyond} to induce seven cognitive load levels in 12 participants. 32 channel EEG signals were recorded at a sampling rate of 256 Hz which were then downsampled to 64 Hz. Wavelet-based features such as entropy, energy, and standard deviation were subsequently extracted from the data and used to train a Multilayer Perceptron (MLP) to classify cognitive load.

In another work \cite{liu2023fusion}, EEG-based functional connectivity, microstates, and PSD features from EEG were used to predict three levels of cognitive load using Support Vector Machine (SVM), Linear Discriminant Analysis, k-Nearest Neighbor (kNN), and Random Forest classifiers. The data was collected from 12 adults while they participated in a three-complexity level experiment. Cognitive load in the context of driving has been explored in \cite{wang2023characterisation} where four driving related tasks were performed by 20 participants. The data was collected using 24-channel EEG device at a sampling rate of 256 Hz. Independent Component Analysis was then applied to reduce the muscle and eye movement artifacts from the data, followed by extraction of PSD feature. The classification performance was then measured for deep neural networks and SVM.

In the end, the relation between affect and cognitive load has been explored through the use of deep transfer learning \cite{pulver2023eeg}. In this approach, a transformer network was pre-trained on two EEG-based emotion recognition datasets before being fine-tuned on a downstream cognitive load dataset. The experiments demonstrated high computational transferability between affect and cognitive load evidenced by the improved performance caused by the pre-training process.

\subsection{Affect classification from EEG}

EEG has been extensively used in emotion classification, with numerous datasets dedicated to this task. Some of the most popular datasets include SEED \cite{duan2013differential, zheng2015investigating}, SEED-IV \cite{zheng2018emotionmeter}, and DEAP \cite{koelstra2011deap}. 
Different datasets employ different labeling schemes for emotion states. For instance, SEED categorizes emotions into positive, negative, and neutral, while SEED-IV includes happy, sad, neutral and fear, whereas DEAP employs arousal, valence, and dominance. Many studies have used these datasets for emotion classification using both machine learning as well deep learning models. Here we will discuss some of those works.

In \cite{bazgir2018emotion}, Discrete Wavelet Transform was used to extract features from 4 EEG frequency bands. Additionally, spectral features were also extracted and Principle Component Analysis was applied for dimensionality reduction. For classification on the DEAP dataset, SVM, kNN, and ANN were used in the same paper. In another study  \cite{alhalaseh2020machine},  emotion classification was performed on the DEAP dataset using entropy and Higuchi’s Fractal Dimension features. Classifiers such as naïve Bayes, KNN, CNN, and Decision Tree were then used for emotion recognition.

In \cite{zhang2023spatio}, the proposed model calculated the average point and the spatial distances derived from covariance matrices on the Riemannian manifold. Next, tangent space learning techniques were used to portray the spatial data onto a Euclidean space and the spatial information were learned using fully connected layers. The proposed model also learned the temporal information from the extracted Differential Entropy (DE) and Power Spectral Density (PSD) features. The performance of the model was evaluated on multiple EEG datasets including SEED. A recent study by \cite{wu2024study} used four functional connection features for emotion recognition. The work presented a Laplace matrix to convert the features into partially positive values, followed by a max operator to ensure all transformed features have strictly positive values. This was followed by a Symmetric Positive Definite matrix network and fully connected layers. This method involves learning hierarchical representations of spatial features through the network's layers, capturing complex patterns and relationships in the data.

In \cite{phan2021eeg} a 2D CNN architecture comprising convolutional layers with various kernel sizes was introduced for the classification of both valence and arousal on the DEAP dataset. The approach uses multi-scale kernel convolution blocks which allow the integration of features from both local and global regions of the input data. 
A spatial-temporal graph attention mechanism was combined with a Transformer encoder in \cite{li2023stgate}. This strategy enables the network to learn time-frequency information for individual channels as well as the exploration of the relationship between different channels. Their performance is verified on SEED, SEED-IV, and DREAMER \cite{katsigiannis2017dreamer} datasets. Another recent study \cite{fan2024eeg}, used Graph Convolutional Networks for emotion classification with EEG. Their proposed network RGNet used a region-wise encoder to extract features from specific channels. The information was then utilized to implement a Graph Convolutional Network. 

To solve the problem associated with unlabeled EEG data, a semi-supervised approach was proposed in \cite{zhang2022parse}. First, data was augmented by applying Gaussian noise to the unlabeled EEG data. Then the unlabeled data were combined to obtain a large dataset and a classifier was used to generate pseudo-labels. A regularization technique called mixup \cite{zhang2017mixup} was then implemented to train the network by combining pairs of data and their labels using convex combinations. Finally, the distribution of embeddings from pairs of labeled and unlabeled data were adjusted to ensure alignment. The performance of the proposed method was tested on multiple EEG emotion datasets. In another work, six different types of transformed EEG signals were generated with corresponding labels to pre-train the self-supervised network in \cite{wang2023self}. The fully-connected layers were then fine-tuned for emotion recognition. Lastly, another self-supervised framework by \cite{zhang2022ganser} leveraged data augmentation and merged adversarial with self-supervised training. The work combined multiple components such as masking transformation operations, UNet network, channel masking operations, and an adversarial network. By combining these elements, the proposed adversarial augmentation network produced synthetic EEG data that closely resembled real EEG recordings, offering increased diversity and quality.

\begin{figure*}[t]
\centering
\includegraphics[width=1\linewidth]{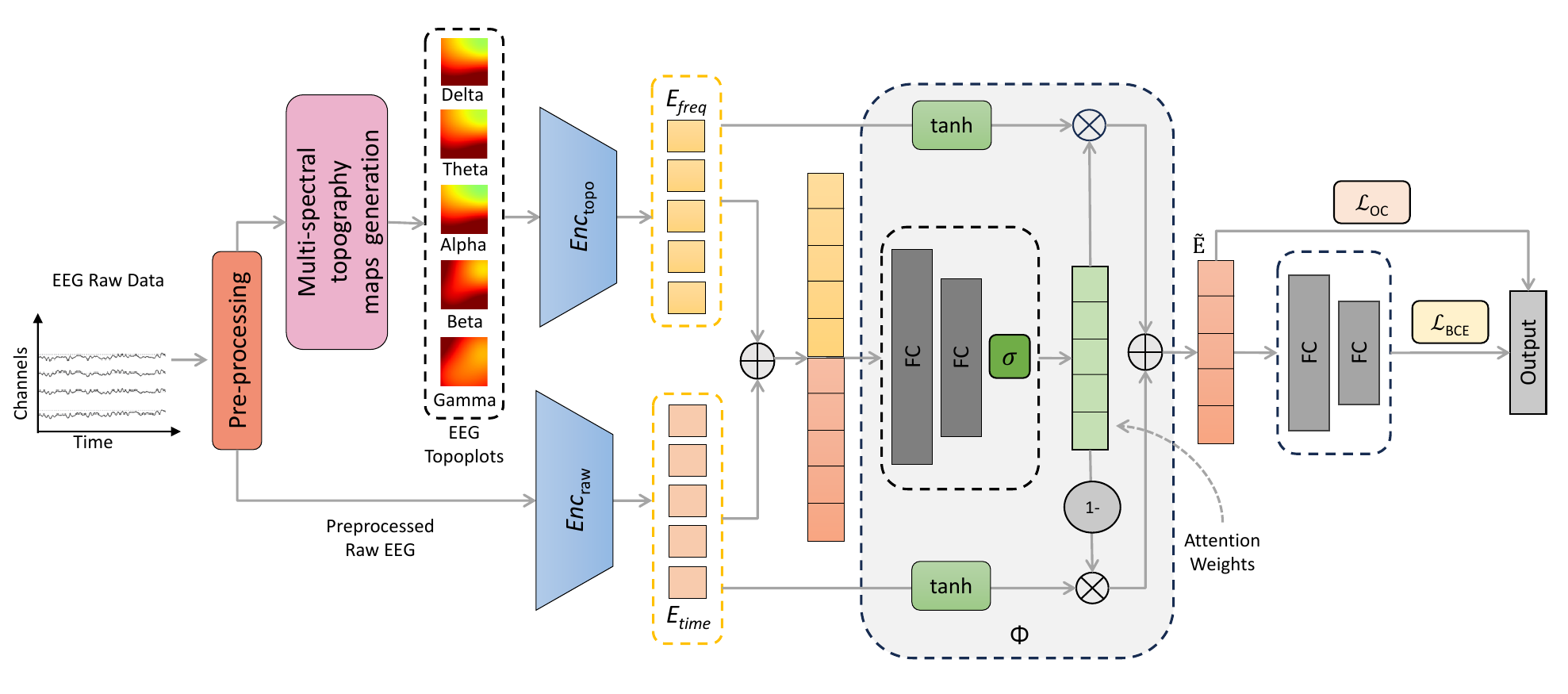}
\caption{The overview of our proposed network. The diagram shows the multi-domain representation learning along with the attention module and the orthogonality constraint.}
\label{model}
\end{figure*}

\section{Method}\label{Method}

\subsection{Problem statement}
Given a set of EEG recordings, $X=[X_1, X_2, \cdots, X_c] \in \mathbb{R}^{c}$ where $c$ is the number of channels, our goal is to extract meaningful information from both time and frequency domains to perform effective cognitive load classification. Let us assume we extract representations from $X$ in time and frequency domains, denoted by $E_\text{\textit{time}} \in \mathbb{R}^{M_1}$ and $E_\text{\textit{freq}} \in \mathbb{R}^{M_2}$, respectively. Prior works have demonstrated that representations from different modalities or domains can often contain diverging distributions and statistical properties. As a result, standard fusion through naive techniques such as $[E_\text{\textit{time}}, E_\text{\textit{freq}}]$ (simple concatenation) may result in an embedding space that is not unified, thus yielding sub-optimal results. The goal of this work is to develop a neural pipeline where we obtain the final fused embedding $\Phi([\tilde{E}_t,\tilde{E}_f])$, where $\tilde{E}_t$, $\tilde{E}_f$, and $\Phi$ are the new embeddings and a transform function respectively, which shows high inter-class distance and enhanced intra-class clustering characteristics.

\subsection{Our approach}
\subsubsection{Multi-domain encoding}

Given an input EEG time-series segment $X$, and our goal of learning discriminate information in both time and frequency domains, we first pass the raw EEG time-series through a convolutional encoder with 1D convolutions, which we refer to as $Enc_\text{raw}$ (see Figure \ref{model}). This yields a set of representations denoted by $E_\text{\textit{time}} \in \mathbb{R}^M$. 

Next, we measure the PSD of all $c$ channels from the raw EEG, followed by dividing the PSD for each channel into the five frequency bands of Delta, Theta, Alpha, Beta, and Gamma. The power of each frequency band is then calculated for all four channels using Simpson's rule \cite{velleman2005generalized}. Subsequently, we use the locations of the four EEG channels to generate the channel power values as 2D images referred to as multi-spectral topography maps \cite{gilberet2017automated, havugimana2023deep}, using the radial basis function interpolator per frequency band. This gives us a $w \times h \times d$ dimension image per frequency band that contains all four channel information. The five frequency band images are then concatenated along the channel axis to generate one image of dimension $w \times h \times D$, where $D = 5d$ is the concatenated channels' information for all five frequency bands. These multi-spectral topography maps are then fed to a separate encoder named $Enc_\text{topo}$ to obtain output embeddings $E_\text{\textit{freq}} \in \mathbb{R}^M$. 

\subsubsection{Embedding fusion}
After separately obtaining the time and frequency embeddings $E_\text{\textit{time}}$ and $E_\text{\textit{freq}}$, we fuse them to generate enriched embeddings by using an attention module $\Phi$. This module, based on \cite{arevalo2017gated,chen2020multi,tu2023dctm,keisham2022online}, maps the domain-specific embeddings onto a shared embedding space to emphasize more on the salient representations with respect to final goal of cognitive load classification. Figure \ref{model} demonstrates the attention mechanism $\Phi$ in detail. The attention scores applied on the concatenated embeddings $[E_\text{\textit{time}}, E_\text{\textit{freq}}]$ are calculated as:
\begin{equation}
    a = \sigma(h_{\text{att}}([E_\text{\textit{time}}, E_\text{\textit{freq}}])),
\end{equation}
where $\sigma$ denotes the sigmoid function, and $h_{\text{att}}$ represents the FC layers in the attention module. Following this, we integrate $E_\text{\textit{time}}$ and $E_\text{\textit{freq}}$ by the attention scores $a \in \mathbb{R}^M$ to generate the merged embeddings $\Tilde{E}$ as:
\begin{equation}
    \Tilde{E} = a \odot  \tanh(E_\text{\textit{freq}}) + (\mathds{1}_{1\times M} - a) \odot  \tanh(E_\text{\textit{time}}),
\end{equation}
where $\odot$ symbolizes element-wise multiplication.

\subsubsection{Orthogonal mapping}
Following obtaining the fused shared embedding, to enhance intra-class clustering while maximizing inter-class separability, we impose an orthogonality constraint on  $\Tilde{E}$. Orthogonality constraints have been applied in prior works such as \cite{jalali2019atrial,ranasinghe2021orthogonal, wang2023shared} to improve learned representations in various uni-modal and multi-modal settings. We apply this notion through applying a loss term that uses cosine similarity as a measure to adjust the alignment and distance between vectors within a mini-batch. Accordingly, the orthogonality constraint loss $\mathcal{L}_{OC}$ is calculated as follows:
\begin{equation}
\label{eq:op-loss}
 \mathcal{L}_{OC} = 1 - \sum_{\substack{i,j \in B \\ y_i = y_j}} \cos(\Tilde{E}_i, \Tilde{E}_j) + \sum_{\substack{i,k \in B \\ y_i \neq y_k}} \cos(\Tilde{E}_i, \Tilde{E}_k),
\end{equation}
where $B$ represents the set of all pairs $(i, j)$ within a mini-batch, and $\Tilde{E}_i$ and $\Tilde{E}_j$ are the embedding vectors corresponding to the indices $i$ and $j$, respectively. $y_i$ and $y_j$ denote the class labels associated with $\Tilde{E}_i$ and $\Tilde{E}_j$, indicating whether they belong to the same class ($y_i = y_j$) or different classes ($y_i \neq y_k$). The cosine similarity between vectors $\Tilde{E}_i$ and $\Tilde{E}_j$ is represented by $\cos(\Tilde{E}_i, \Tilde{E}_j)$ aiming to measure the closeness or alignment of these vectors in the feature space. $\mathcal{L}_{OC}$ aims to enhance within-class compactness and between-class separability. The loss function minimizes the cosine similarity between embeddings of different classes (to enforce orthogonality), and maximizes the cosine similarity within embeddings of the same class. 

The cosine similarity is defined as $\cos(u, v) = \frac{u \circ v}{||u|| ||v||}$ in which $||u||$ and $||v||$ are the magnitudes (or Euclidean norms) of vectors $u$ and $v$, and $\theta$ is the angle between them. By normalizing this dot product with their norms, we focus purely on the directional similarity, regardless of the magnitude. Therefore, Eq. \ref{eq:op-loss} can be represented as
\begin{align}
\label{eq:op-loss2}
\mathcal{L}_{OC} = &\ 1 - \sum_{\{i,j \;|\; B; y_i = y_j\}} \frac{\Tilde{E}_i \circ \Tilde{E}_j}{||\Tilde{E}_i|| ||\Tilde{E}_j||} \nonumber \\
&+ \sum_{\{i,k \;|\; B; y_i \neq y_k\}} \frac{\Tilde{E}_i \circ \Tilde{E}_k}{||\Tilde{E}_i|| ||\Tilde{E}_k||}.
\end{align}

For vectors of the same class, we aim to reduce the angle towards 0 degrees, implying a similarity of 1. However, for vectors of different classes, we intend to increase the angle towards 90 degrees which implies a similarity towards 0, enhancing separation. A depiction of the impact of the orthogonality constraint was depicted earlier in Figure \ref{fig:oc_loss}.
\textcolor{black}{For each mini-batch during training, we compute the orthogonal loss by forming the feature covariance (Gram) matrix of that batch’s features, and adding a penalty for deviation from the identity matrix (as in Eq. \ref{eq:op-loss2}). This loss is computed separately for each batch, so variations between batches are naturally handled through the batch-wise computation.}

\subsubsection{Total loss}
In addition $\mathcal{L}_{OC}$ we also apply the standard cross-entropy loss during training. Accordingly, we define the total loss $\mathcal{L}_\textit{\text{total}}$:
\begin{equation}
 \mathcal{L}_\textit{\text{total}} = \mathcal{L}_{CE} + \beta \mathcal{L}_{OC},
 \label{total_loss}
\end{equation}
where $\beta$ is the parameter controlling the influence of $L_{OC}$ within the overall loss calculation.

\subsection{Data Preparation}
\textcolor{black}{For preparing the data, we follow a similar pre-processing method as \cite{pulver2023eeg}. Specifically, to reduce undesired noise and artifacts, we first apply a 2nd order Butterworth bandpass filter with a passband frequency ranging from 1 to 75 Hz. Furthermore, a notch filter with a quality factor of 30 is employed to reduce power line noise at a frequency of 60 Hz. We then segment the data into 10-second segments for 5 frequency bands, namely Delta (1-4 Hz), Theta (4-8 Hz), Alpha (8-12 Hz), Beta (12-31 Hz), and Gamma (31-75 Hz). Subsequently, we extract the widely used absolute PSD features \cite{zhang2023distilling, pulver2023eeg, angkan2024multimodal} for each segment and for each frequency band. PSD serves as an indicator of the signal power distributed across various frequency bands with no overlap. To compute this feature, we employ Welch's method \cite{solomon1991psd}. Welch's method involves dividing the signal into smaller sections, then calculating the discrete Fourier Transform for each segment, followed by averaging their squared magnitudes which offers smoother and more reliable spectral estimates by averaging over segments. Finally, we apply z-score normalization \cite{dai2021electroencephalogram} to the extracted PSD features. In CL-Drive data, the absolute PSD values before z-score normalization ranges approximately from 0.55 to 55598.56  $\mu$V$^2$ and for CLARE it ranges from 0.67 to 301912.90 $\mu$V$^2$. Rather than performing a time-frequency analysis like STFT, we focused on the overall spectral content per segment. This approach aligns with cognitive load analysis, where sustained frequency power in different bands is more informative than rapid temporal changes \cite{kathner2014effects,pergher2019mental}. We then projected the PSD values of the 4 channels onto a 2D plane using electrode positions based on the 10-20 system and interpolated them into $32 \times 32$ topographic images using RBF interpolation. Each of the five frequency bands produced one such image, color-mapped via the Jet colormap, and concatenated to form a $32 \times 32 \times 15$ tensor used as input to the encoder.}

\subsection{Multi-spectral topography maps}
\textcolor{black}{To generate the multi-spectral topography matrices used in our pipeline, we first generate the PSD for each channel using 10-second windows as discussed above. The PSD plots are then segmented to the five frequency bands mentioned earlier and each band's power is then calculated. Next, we find the electrode locations according to the international 10-20 system \cite{silverman1963rationale} for the four electrodes of the EEG headband. We then interpolate the $4 \time 1$ vector for each band into a $32 \times 32 \times 3$ image using the radial basis function (RBF) interpolator from the SciPy library\footnote{https://docs.scipy.org/doc/scipy/reference/generated/scipy.interpolate.Rbf.html}. The dimension that represents the three color channels is generated based on the Jet colormap. For each topographic map generated, we use a symmetric color bar centered around zero to represent the spatial distribution of EEG power values. Specifically, we compute the maximum absolute value from the interpolated topographic data and set this as the upper bound (vmax) of the color scale. The lower bound (vmin) is set to the negative of this value, i.e., [-vmax, +vmax]. This approach ensures a balanced visualization where positive and negative deviations from the mean are equally represented in the colormap. The use of symmetric limits also avoids misleading interpretations caused by shifting color scales across different trials. The output images for each band are henceforth referred to as topography maps. We then concatenate the topography maps for the 5 frequency bands (referred to as spectrum) to create a $32 \times 32 \times 15$ matrix to feed $Enc_\text{topo}$.}

\subsection{Implementation Details}\label{Implementation Details}
\textcolor{black}{Here, we provide a detailed description of our implemented pipeline. For all the training sessions, we use a batch size of 32 and the Adam optimizer \cite{adam} with a learning rate 0.0001. A learning rate scheduler Plateau was used with a decay ratio of 0.1 and patience of 10. The training is done for 200 epochs. Training was performed on an NVIDIA 2080 Ti GPU, and our pipeline was implemented using the PyTorch library.}

\textcolor{black}{Both encoders, $Enc_\text{raw}$ and $Enc_\text{topo}$, consist of 3 convolutional blocks, each containing two convolution layers with batch norm and ReLU. The attention mechanism consists of a two-layer fully-connected network with 512 and 256 neurons, respectively. The size of $E_\text{\textit{time}}$ and $E_\text{\textit{freq}}$ are both set to  256, while the size of $\Tilde{E}$ is also 256. Finally, the fully connected layers in the classification block are of size 256 and 128, respectively.
The details of all the components of our pipeline are presented in Table \ref{table:model_details}.}

\section{Experiment setup}\label{Experimentation}
In this section, we explain the specifics of the experiments conducted in this study. Initially, we explain the datasets and data preparation process. Then we outline the protocol utilized to evaluate the effectiveness of our approach. Next, we elaborate on the implementation details necessary for reproducing our experiments.

\subsection{Datasets} \label{Datasets}
\begin{figure}[t]
\centering
\includegraphics[width=0.5\columnwidth]{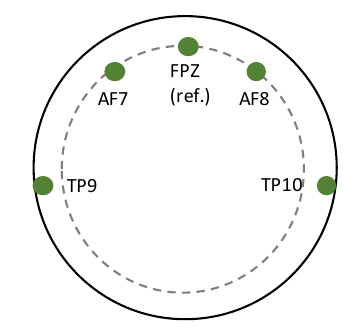}
\caption{EEG electrode locations according to the international 10-20 system.}
\label{electrodes}
\end{figure}

We use two public EEG-based cognitive load datasets, CL-Drive \cite{angkan2024multimodal} and CLARE \cite{bhatti2024clare}. Both datasets contain four types of physiological signals, among which, we explicitly use EEG. The EEG signals have been collected with a Muse S\footnote{https://choosemuse.com/} headband which uses 4 channels and a sampling frequency of 256 Hz. The EEG signals in both datasets have been recorded following the international 10-20 electrode placement system using 4 electrodes, namely TP9, AF7, AF8, and TP10. The reference electrode is located at Fpz, which is at the middle of the forehead. The AF7 and AF8 electrodes sense the prefrontal cortex of the brain, while TP9 and TP10 sense the temporal region of the brain. The prefrontal cortex is directly affected by cognitive load, whereas the temporal area is involved in processing auditory information, memory encoding, and affect. Since cognitive load is influenced by affect, and both datasets in this study involve auditory information through driving tasks or tasks involved in the MatB-II, all these electrodes play a vital role in capturing changes in brain activities. Figure \ref{electrodes} depicts the sensor placement for the EEG device used to collect both datasets. Following we briefly describe each dataset in more detail.

\subsubsection{CL-Drive}
The CL-Drive data have been collected from 23 participants (17 female and 6 male) while they drove a vehicle simulator and completed 9 tasks of different complexity levels for 3 minutes each. Out of the 23 participants, sessions for two individuals were stopped due to SAS \cite{angkan2024multimodal}, resulting in a dataset consisting of data from 21 participants.

\subsubsection{CLARE}
The CLARE dataset has been collected from 24 participants, of which data from 19 participants are made publicly accessible. The data were collected while the participants performed one or multiple tasks in the Multi-Attribute Task Battery-II (MATB-II) \cite{santiago2011multi} software developed by NASA\footnote{https://www.nasa.gov/}, which can induce varying levels of cognitive load.

\subsubsection{Cognitive load labels} 
In both CL-Drive and CLARE datasets, self-reported cognitive load scores have been collected every 10 seconds during the experiment using the Paas scale \cite{paas1992training} which is a 9-point Likert scale where 1 represents ``very, very low mental effort'' and 9 indicates ``very, very high mental effort'' experienced by the subject. The Paas scale is the most commonly used cognitive load self-reporting tool and has been used to measure different types of cognitive load experienced by participants in various studies \cite{paas1994variability, sweller2018measuring, solhjoo2019heart, ayres2006using, asaadi2024effects}. The cognitive load scores were then transformed into binary values, with a range of 1 to 5 representing low cognitive load and 6 to 9 indicating high cognitive load. 

\subsection{Baselines}
We compare our proposed model with several baselines from recent works in the area. Specifically, both CL-Drive and CLARE have a number of baseline results based on CNN architectures, which we employ. To obtain the best results possible, we optimize the parameters and re-run the experiments based on the proposed architectures, i.e., VGG (with extracted PSD features), ResNet (with extracted PSD features), VGG (with raw data), and ResNet (with raw data) \cite{angkan2024multimodal}. Both the VGG-style and ResNet-style networks, when trained on extracted features, consist of two conv1D blocks followed by a classification block as proposed in \cite{angkan2024multimodal}. When these networks are trained on raw data, they include three conv1D blocks followed by a classification block \cite{angkan2024multimodal}. Additionally, we use the popular Conformer \cite{song2022eeg} model as a baseline. This network is trained on raw data and consists of 10 attention heads. The depth of the network is 6 and the temporal pooling filter length is 75 with a stride of 15. We also use the Masked Autoencoder (MAE) model proposed in \cite{pulver2023eeg} as a baseline using PSD features. This model consists of 4 attention blocks and a prediction head with 2 linear layers followed by a classification layer. 
We also experiment and compare our method with the  Multi-Modal Cosine (MMCosine) loss as stated in \cite{xu2023mmcosine}. The MMCosine loss enhances the similarity of intra-class features while making them more distinguishable from those of other classes. This is achieved by maximizing the cosine similarity between samples of the same class. As a result, the angular distance between vectors of the same class is reduced, enabling the model to more effectively identify the inter-class differences. In our experiment, the MMCosine loss is applied to the feature embeddings $Enc_\text{freq}$ and $Enc_\text{time}$.

\textcolor{black}{Additionally, we used Filter Bank Common Spatial Pattern-Support Vector Machine (FBCSP-SVM) \cite{ang2008filter,sundaresan2021evaluating} and Common Spatial Pattern-Support Vector Machine (CSP-SVM) \cite{ramoser2000optimal,luo2021feature}. 
For our FBCSP-SVM implementation, segmented EEG raw data are partitioned in the frequency domain using a filter bank comprising five bandpass filters: 1–4 Hz, 4–8 Hz, 8–12 Hz, 12–31 Hz, and 31–75 Hz. For each band, the common spatial pattern (CSP) algorithm was applied to extract spatial filters that maximize the variance difference between the two classes. We selected the top 2 and bottom 2 eigenvectors per band (4 CSP components per band), resulting in a total of 20 features per trial. These CSP features were concatenated and used to train a SVM classifier with a RBF kernel. We used the default 'scale' setting for gamma and a regularization parameter C = 1.0.
}
\textcolor{black}{For the CSP-SVM model, each EEG sample was bandpass filtered between 1–75 Hz to capture a broad range of frequency information relevant to cognitive load. Feature extraction was performed using the Common Spatial Pattern (CSP) algorithm, which computes spatial filters that maximize variance differences between the two classes. We used 4 CSP components per class (total of 8 features) to construct the feature representation for classification. Similarly SVM with a RBF kernel was trained on these features, using a regularization parameter C=1.0 and kernel coefficient $\gamma$ is set to 0.1. All implementations were carried out in Python using the scikit-learn library.}


\subsection{Evaluation Protocol}
\textcolor{black}{For evaluating our proposed pipeline, we follow the binary Leave-One-Subject-Out (LOSO) evaluation scheme proposed in the original CL-Drive \cite{angkan2024multimodal}  and CLARE \cite{bhatti2024clare} dataset papers. We report both the accuracy and the F1 scores in this paper since the class distribution of the dataset is not perfectly balanced.}

\begin{table}
\centering
\caption{Architectural details of our proposed network used in this study for both raw data and multi-spectral topography maps}.
\label{table:model_details}
\resizebox{1\columnwidth}{!}{
\begin{tabular}{l|l|l|l}
 \hline
 \textbf{Modules} & \textbf{Parameters} &  \textbf{\makecell{$Enc_\text{raw}$}} & \textbf{\makecell{$Enc_\text{topo}$}} \\ 
 \hline
 \hline
 Conv Block 1  & Layer type & Conv1D & Conv2D \\
 & No. of Layers & 2 & 2 \\
 & Activation & ReLU & ReLU \\
 & Kernel size & $1 \times 32$ & $3 \times 3$\\
 & Input channel & 4 & 15 \\
 & Output channel & 64 & 64 \\
 & Padding & 0 & Same \\
 & MaxPool & 1D & 2D \\
 \hline
 Conv Block 2 & Layer type & Conv1D & Conv2D \\
 & No. of Layers & 2 & 2 \\
 & Activation & ReLU & ReLU \\
 & Kernel size & $1 \times 16$ & $3 \times 3$\\
 & Input channel & 64 & 64 \\
 & Output channel & 128 & 128 \\
 & Padding & 0 & Same \\
 & MaxPool & 1D & 2D \\
 \hline
 Conv Block 3 & Layer type & Conv1D & Conv2D \\
 & No. of Layers & 2 & 2 \\
 & Activation & ReLU & ReLU \\
 & Kernel size & $1 \times 8$ & $3 \times 3$\\
 & Input channel & 128 & 128 \\
 & Output channel & 256 & 256 \\
 & Padding & 0 & Same \\
 & MaxPool & 1D & 2D \\
 \hline
 Attention Mechanism & Layer type & \multicolumn{2}{c}{FC}\\
  & Number of layers & \multicolumn{2}{c}{2} \\
 & Layer shape & \multicolumn{2}{c}{$512 \times 256$}\\
 & Activation & \multicolumn{2}{c}{ReLU} \\
 \hline
 Classification & Layer type & \multicolumn{2}{c}{FC}\\
 Block & Number of layers & \multicolumn{2}{c}{2} \\
 & Layer shape & \multicolumn{2}{c}{$256 \times 128$}\\
 & Dropout rate & \multicolumn{2}{c}{0.5}\\
 & Activation & \multicolumn{2}{c}{ReLU} \\
 \hline
 \end{tabular}
}
\end{table}

\begin{table*}[t]
    \caption{Performance of our proposed method compared to several state-of-the-art solutions.}
    \label{table:comparison}
    \small
    \setlength
    \tabcolsep{8pt}
        \begin{center}{
            \begin{tabular}{l|ll|ll}
                 \hline
            \multicolumn{1}{c|}{} &
            \multicolumn{2}{c|}{CL-Drive}& \multicolumn{2}{c}{CLARE}\\
                 Model   & Accuracy & F1 score &  Accuracy & F1 score\\
                 \hline\hline
                 \textcolor{black}{FBCSP-SVM} \cite{ang2008filter} & 59.46(22.63) & 49.41(22.31) & 61.14(21.43) & 41.17(16.54)\\
                 \textcolor{black}{CSP-SVM} \cite{luo2021feature} & 57.65(22.44) & 51.19(20.59) & 56.61(24.60) & 47.83(20.16)\\
                 VGG-style \cite{angkan2024multimodal} & 70.47(13.37) & 59.52(7.06) & 61.83(14.56) & 54.64(10.63)\\
                 ResNet-style \cite{angkan2024multimodal} & 70.04(16.80) & 60.1(14.48) & 55.82(12.67) & 48.96(10.93)\\
                 VGG-style (rawdata) \cite{angkan2024multimodal} & 70.28(10.87) & 63.12(9.39) & 70.29(16.03) & 60.24(13.16) \\
                 ResNet-style (rawdata) \cite{angkan2024multimodal} & 70.64(9.93) & 59.90(8.54) & 70.87(19.08) & 55.25(15.38) \\
                 Conformer \cite{song2022eeg} & 69.38(8.72) & 63.29(9.29) & 70.42(16.02) & 58.28(12.00)\\ 
                 MAE \cite{pulver2023eeg}  & 67.88(14.67) & 61.25(13.18) & 62.48(10.71) & 57.51(7.29) \\
                 MMCosine \cite{xu2023mmcosine} & 73.64(7.74) & 64.84(7.99) & 71.48(16.27) & 58.35(13.67) \\
                 \textbf{Proposed}    & \textbf{75.40(7.15)} & \textbf{66.00(7.78)} & \textbf{74.96(15.51)} & \textbf{61.62(13.74)}   \\
                 
                 \hline
                \end{tabular} 
                }
        \end{center}
\end{table*}

\begin{table*}[t]
    \caption{Ablation experiments demonstrating the impact of each module within our proposed model. MS in this table stands for multi-spectral.}
    \label{table:ablation}
    \small
    \setlength
    \tabcolsep{8pt}
        \begin{center}{
            \begin{tabular}{cccc|ll|ll}
                 \hline
             &      &      &\multicolumn{1}{c|}{} &
            \multicolumn{2}{c|}{CL-Drive}& \multicolumn{2}{c}{CLARE}\\
                 Rawdata    & MS topography maps &$\mathcal{L}_{OC}$ &Attention ($\Phi$) &Accuracy & F1 score &  Accuracy & F1 score\\
                 \hline\hline
                 \cmark &\cmark  &\cmark  &\cmark & \textbf{75.40(7.15)} & \textbf{66.00(7.78)} & \textbf{74.96(15.51)} & \textbf{61.62(13.74)}\\
                 \cmark &\cmark  &\xmark  &\cmark & 71.20(10.21) & 63.55(9.03) & 68.72(16.80) & 58.58(14.24)  \\
                 \cmark &\cmark  &\cmark  &\xmark & 71.06(6.39) & 63.68(6.32) & 71.93(14.50) & 59.73(11.86)\\
                 \cmark &\cmark  &\xmark  &\xmark & 71.30(9.42)& 62.27(7.00) & 69.58(16.71) & 55.58(13.83)\\
                 \cmark &\xmark  &\xmark  &\xmark & 70.19(7.26) & 63.44(6.58) & 70.46(16.99) & 57.76(13.69)\\
                 \xmark &\cmark  &\xmark  &\xmark & 68.17(11.97) & 60.27(9.23) & 69.78(17.90) & 55.46(14.73)\\
                 \hline
                \end{tabular} 
                }
        \end{center}
\end{table*}

\begin{table*}[t]
    \caption{The impact of different values of $\beta$ with and without the Attention module.}
    \label{table:sensitivity}
    \small
    \setlength
    \tabcolsep{15pt}
        \begin{center}{
            \begin{tabular}{c|l|ll|ll}
                 \hline
            \multicolumn{1}{c|}{Attention} &\multicolumn{1}{c|}{$\beta$} &
            \multicolumn{2}{c|}{CL-Drive}& \multicolumn{2}{c}{CLARE}\\
                  & & Accuracy & F1 score &  Accuracy & F1 score\\ \hline\hline
                 &0.4 & 71.06(6.39) & 63.68(6.32) & 71.93(14.50) & 59.73(11.86) \\
                 \xmark &0.7 &  71.64(7.85) & 64.36(7.63) & 71.60(16.12) & 58.31(12.18) \\
                 &1 & 71.75(10.12) & 63.63(8.87) & 69.71(17.15) & 60.55(14.99) \\ \hline
                 & 0.4 & \textbf{75.40(7.15)} & \textbf{66.00(7.78)} & \textbf{74.96(15.51)} & \textbf{61.62(13.74)} \\
                 \cmark &0.7 & 71.49(8.53) & 64.77(8.27) & 73.15(16.27) & 60.96(13.48) \\
                 &1 & 72.70(8.98) & 65.18(7.64) & 69.99(19.47) & 56.82(17.46) \\
                 \hline
                \end{tabular} 
                }
        \end{center}
\end{table*}


\section{Results}\label{results}  
\label{Results}
In this section, we evaluate the proposed method for EEG-based cognitive load classification through a series of experiments on the CL-Drive and CLARE datasets. Initially, we compare the efficacy of our model against the state-of-the-art solutions in the literature. Subsequently, we assess the robustness of the model against noise for both datasets. Further analytical studies are conducted through ablation studies, which are complemented by sensitivity analyses focusing on components and parameters within our framework. The numerical accuracy and F1 values are accompanied by their respective standard deviations which are presented in parentheses adjacent to each score. In all tabular representations, the highest results are emphasized in bold.



\subsection{Performance}\label{FrozenVFine}
\begin{figure*}[t]
    \begin{subfigure}[t]{0.32\linewidth}
        \centering
        \includegraphics[width = 1\linewidth]{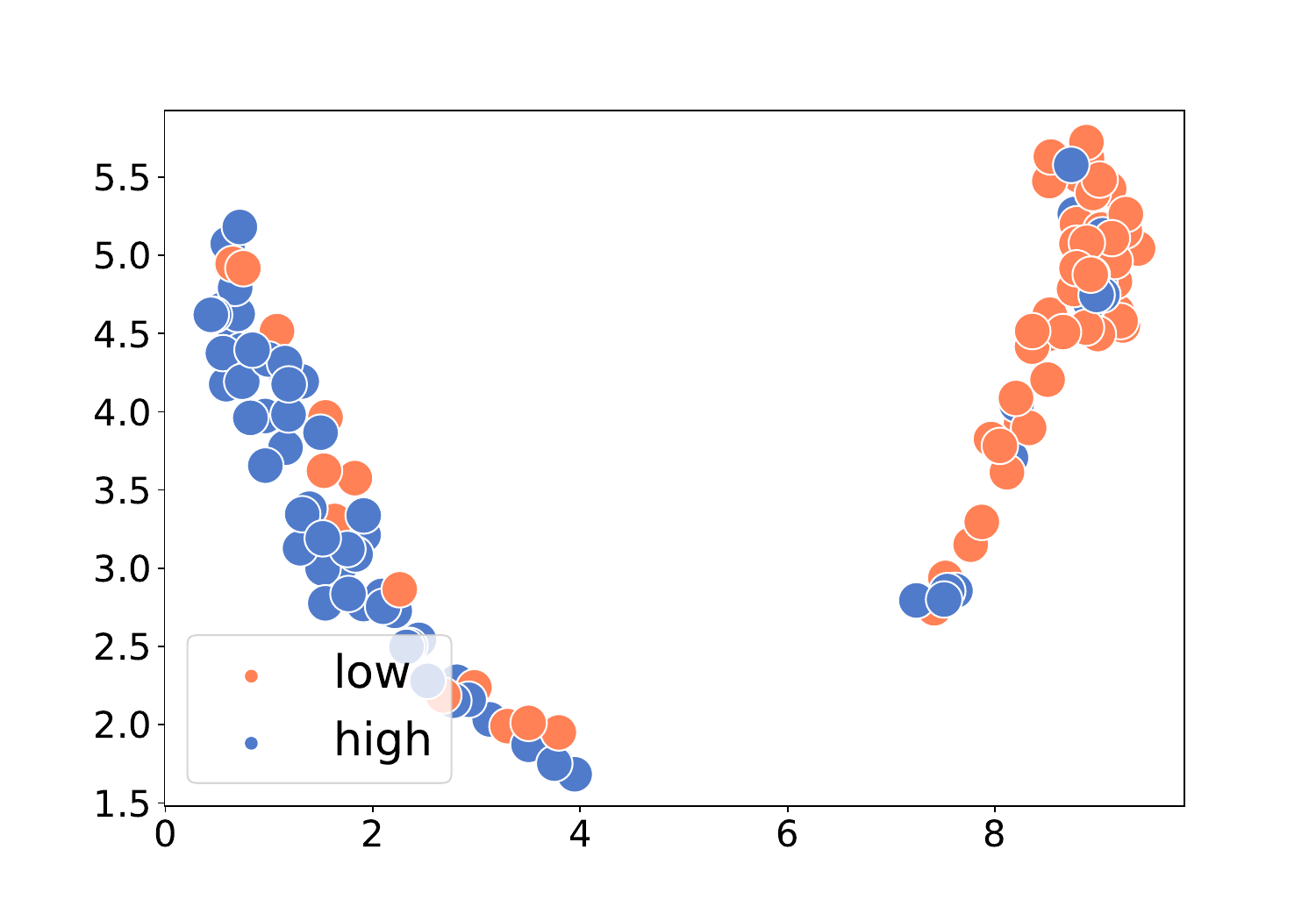}
        \caption{\small Proposed model}
    \end{subfigure}
    \begin{subfigure}[t]{0.32\linewidth}
        \centering
        \includegraphics[width = 1\linewidth]{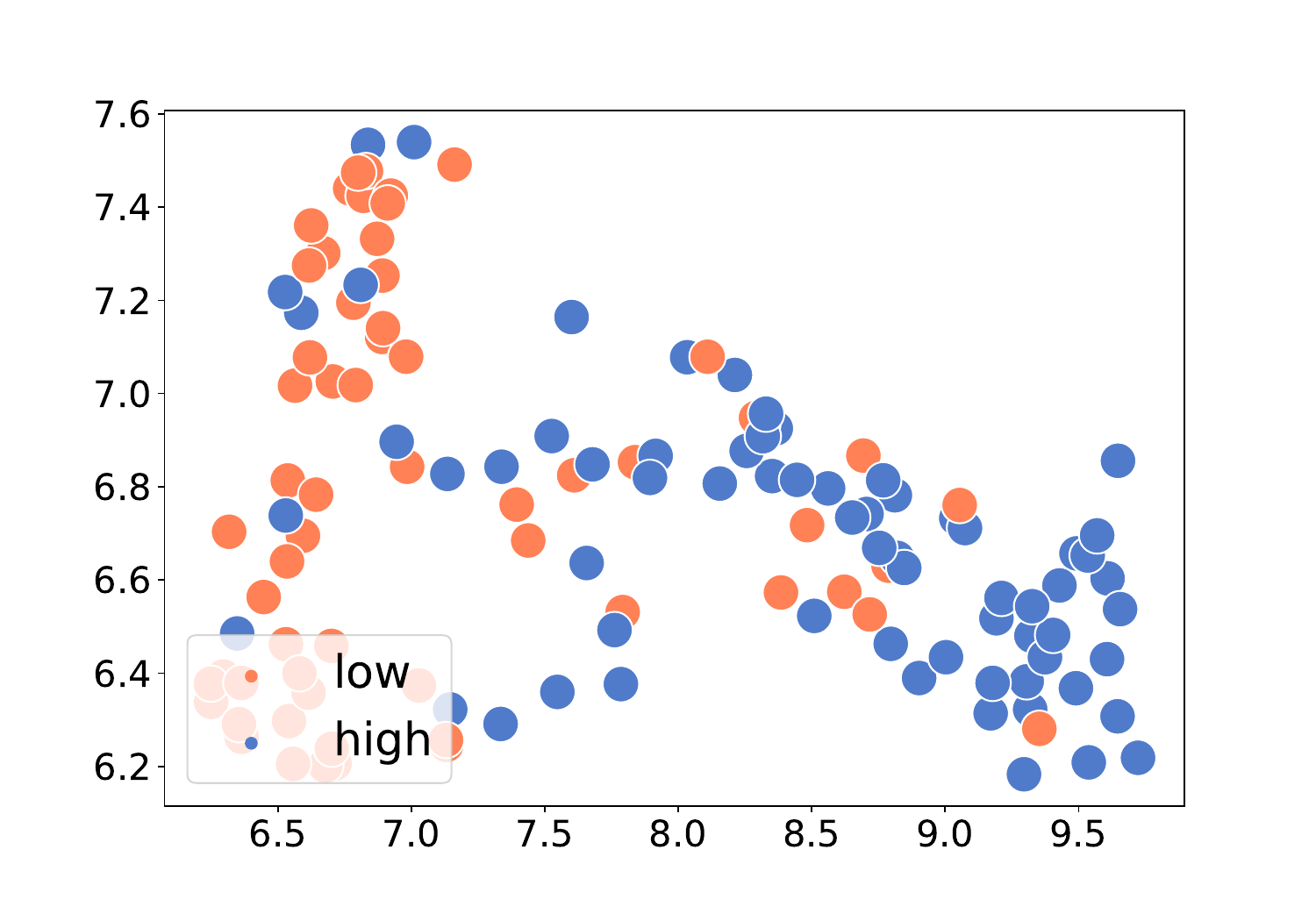}
        \caption{\small W/o \( \mathcal{L}_{OC} \)}
    \end{subfigure}
    \centering
    \begin{subfigure}[t]{0.32\linewidth}
        \centering
        \includegraphics[width = 1\linewidth]{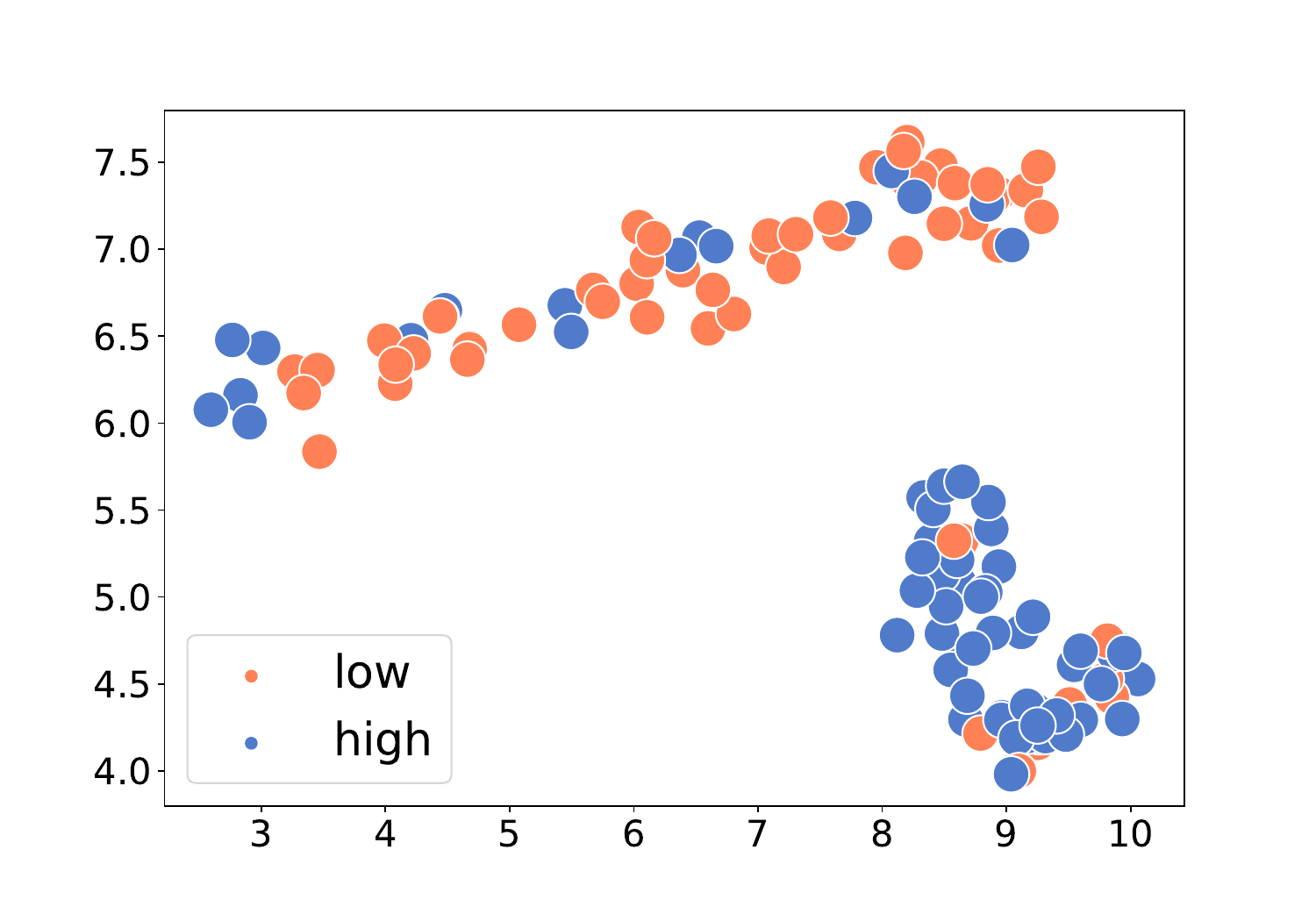}
        \caption{\small W/o attention}
    \end{subfigure}
    \centering
    \begin{subfigure}[t]{0.32\linewidth}
        \centering
        \includegraphics[width = 1\linewidth]{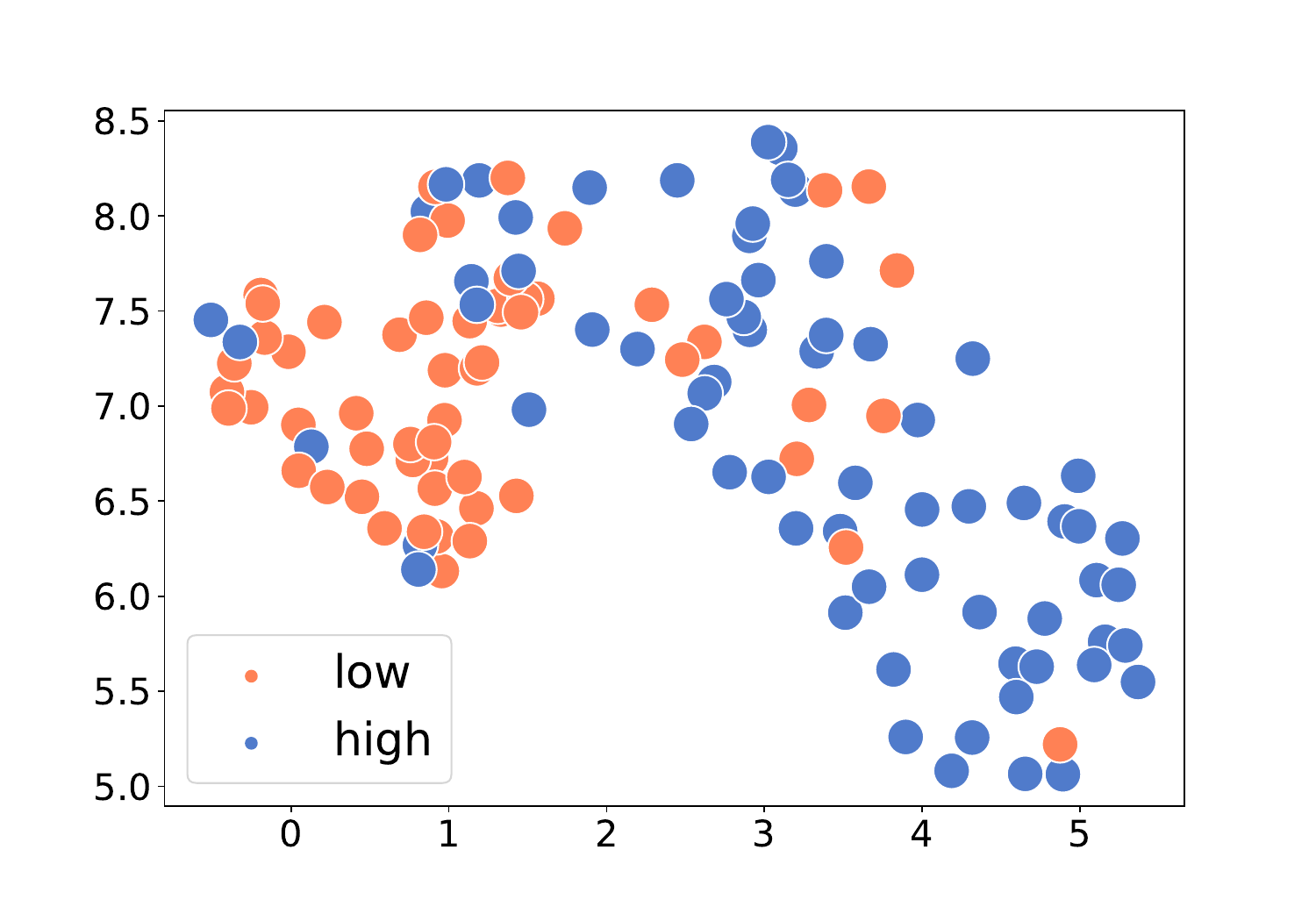}
        \caption{\small W/o \( \mathcal{L}_{OC} \) and attention}
    \end{subfigure}
    \begin{subfigure}[t]{0.32\linewidth}
        \centering       
        \includegraphics[width = 1\linewidth]{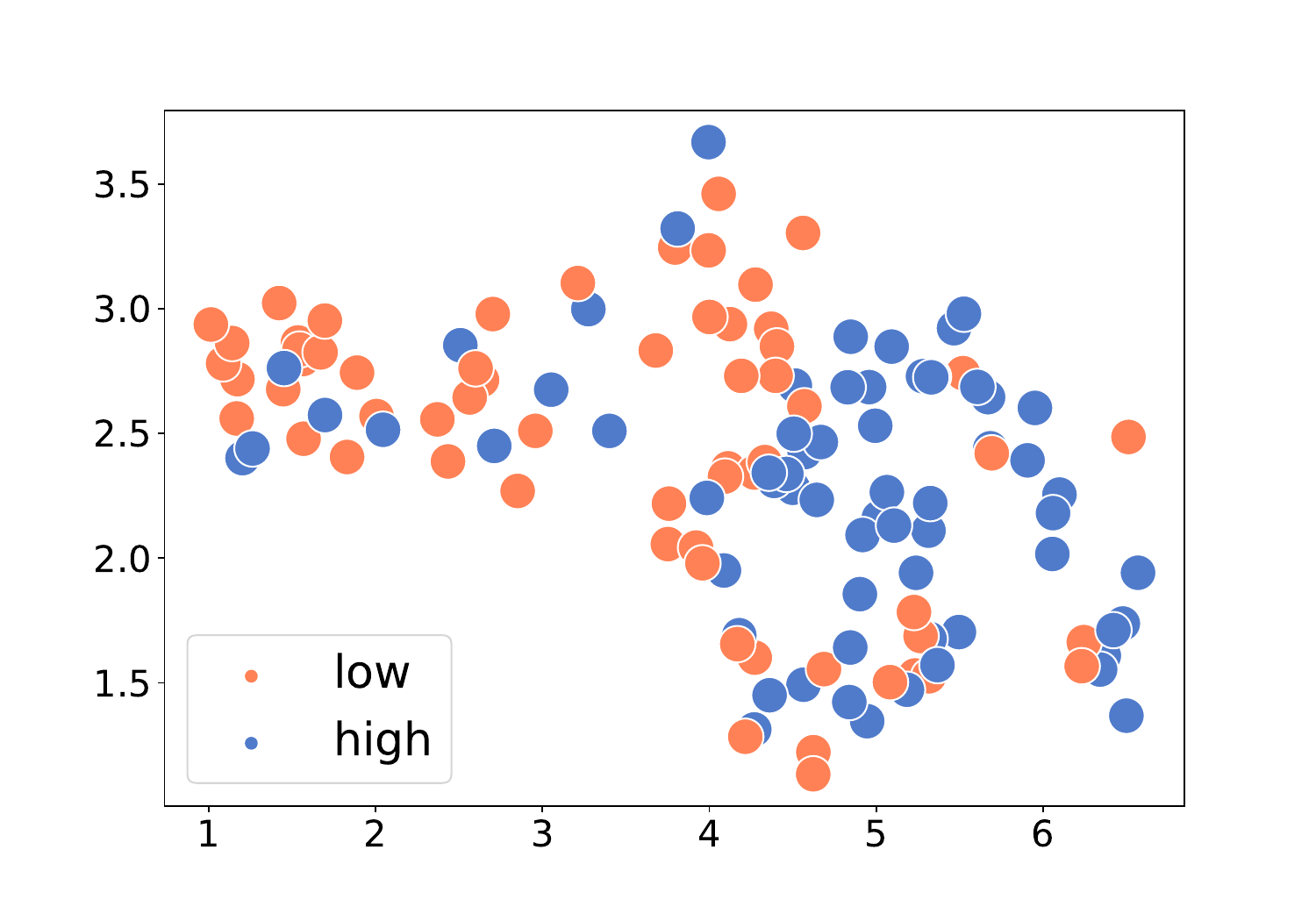}
        \caption{\small W/o multi-spectral topography maps
        }
    \end{subfigure}
    \begin{subfigure}[t]{0.32\linewidth}
        \centering       
        \includegraphics[width = 1\linewidth]{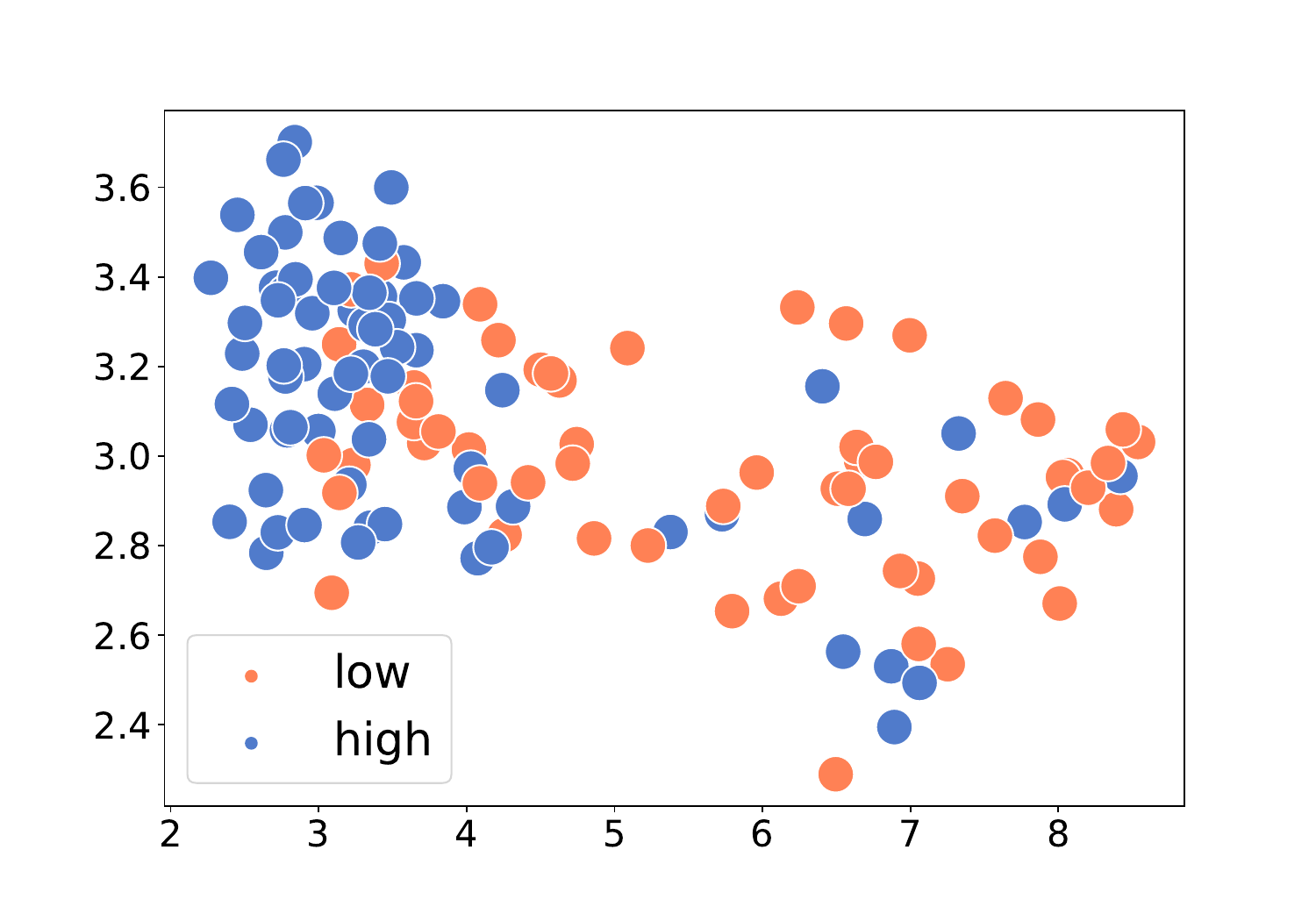}
        \caption{\small W/o raw data
        }
    \end{subfigure}
    \centering
        \centering
    \caption{UMAP visualizations of the CL-Drive dataset under different model configurations.}
\label{fig:umap}
\end{figure*}

We compare the performance of the proposed model with the state-of-the-art methods in literature for cognitive load classification based on EEG. The results of these experiments are summarized in Table \ref{table:comparison}, where we use the multi-domain data (time-series and multi-spectral topography maps).
The results demonstrate higher performance compared to VGG- and ResNet-style models \cite{angkan2024multimodal} that either use the raw EEG data or a combination of features to train the models. \textcolor{black}{Combining time and frequency domains yields substantial improvements (75.40\% accuracy on CL-Drive, 74.96\% on CLARE and 66.00\% F1 score on CL-Drive, 61.62\% on CLARE) over single-domain approaches. The complementary information from raw EEG signals and multi-spectral topography maps creates more distinctive feature representations, as evidenced by the clearer class separation in UMAP visualizations (Figure \ref{fig:umap}).}


\textcolor{black}{The performance gain over Conformer \cite{song2022eeg} (6.02\% and 2.71\% higher accuracy and F1 score on CL-Drive and 4.54\% and 3.34\% higher accuracy and F1 score on CLARE) is particularly noteworthy, as Conformer specifically employs self-attention mechanisms for EEG analysis but lacks our multi-domain integration capability. Similarly, our method surpasses MAE \cite{pulver2023eeg} by 7.52\% and 12.48\% accuracy on CL-Drive, despite MAE's sophisticated architecture.} 

\textcolor{black}{Looking further into the results we observe that for both datasets, the MMCosine \cite{xu2023mmcosine} (PSD) model exhibits the second-best overall performance after our method, when applied to multi-spectral topography maps. This model performs 1.76\% worse in accuracy and 1.16\% worse in F1 than our method on CL-Drive. On CLARE, the accuracy drops by 3.48\% and the F1 score decreases by 3.27\%. FBCSP-SVM shows lowest performance in terms of F1 score in both the datasets. These consistent improvements across different datasets demonstrate that our domain integration strategy more effectively captures the complementary information necessary for robust cognitive load classification. }

\subsection{Multi-spectral topography maps with Differential Entropy}\label{DE_experiments}
We also apply our method using Differential Entropy (DE), another commonly used frequency-domain feature in EEG research. DE quantifies the randomness or uncertainty within a signal, making it a valuable metric in EEG analysis, particularly for evaluating the randomness within specific frequency bands. To compute DE, we first perform a Fast Fourier Transform (FFT) on the EEG data for each channel, transforming the time-domain signals into the frequency domain. The resulting frequency-domain data is then segmented into five distinct frequency bands, and DE is calculated for each band across all channels. 
We then 
compute the DE features for each frequency band from the four EEG channels.
The process of generating these topography maps is similar to the approach used for the multi-spectral topography maps, ensuring consistent representation across different spectral features.

In Table S1 in the supplementary material
we present the results of our proposed method compared to MAE \cite{pulver2023eeg}, ResNet-style \cite{angkan2024multimodal}, VGG-style \cite{angkan2024multimodal}, and MMCosine \cite{xu2023mmcosine} using the DE multi-spectral topography maps. Our method achieves the highest performance in both CL-Drive and CLARE datasets when trained using both the raw data and multi-spectral topography maps using DE features. The second best performance is obtained by MMCosine, which also utilizes both time and frequency domain features. Specifically, MMCosine is 0.35\% lower in accuracy and 1.59\% lower in F1 for the CL-Drive dataset, and 2.79\% lower in accuracy and 0.37\% lower in F1 for the CLARE dataset. Overall, we observe that the frequency domain topography maps generated using the PSD features, as shown in Table \ref{table:comparison}, yield better results compared to the multi-spectral topography maps using DE features.

\subsection{Ablation Studies}\label{Ablation} 
In this section, we conduct ablation studies to assess the individual contributions of each component within our proposed model and to understand their collective impact on the overall performance. Specifically, we investigate the role of the raw EEG data, the use of the multi-spectral topography maps, the impact of $\mathcal{L}_{OC}$, and the influence of the Attention module. The results of these experiments are presented in Table \ref{table:ablation}, which illustrate a significant performance improvement when all components are integrated concurrently. First, we explore the impact of removing the orthogonal loss, $\mathcal{L}_{OC}$, revealing a substantial decline in performance, which highlights its critical role. Specifically, in Table \ref{table:ablation} a decrease of 4.2\% in accuracy and 2.45\% decrease in F1 are observed for CL-Drive dataset and a decrease of 6.24\% in accuracy and 3.04\% decrease in F1 are noticed for CLARE. Additionally, when the Attention module is removed from our method, the accuracy and F1 decrease by 4.34\% and 2.32\% for CL-Drive, while decreasing by 3.03\% and 1.89\% for CLARE datasets respectively. Moreover, when we remove both $\mathcal{L}_{OC}$ and Attention module and concatenate the time-series and multi-spectral topography maps representations directly, we notice a decrease of 4.1\% and 3.73\% in accuracy and F1 score for CL-Drive dataset, while for CLARE the accuracy and F1 decrease by 5.38\% and 6.04\%. Finally, ablated variations of our approach where only a single domain is used exhibit considerable declines in performance, demonstrating the positive impact of learning the data in both time and frequency domains in conjunction with $\mathcal{L}_{OC}$ and Attention. Note that when an entire domain is ablated, $\mathcal{L}_{OC}$  and attention are also removed as their primary role is to improve the fusion of the information from the two domains.

Figure \ref{fig:umap} presents UMAP visualizations of the CL-Drive dataset under different model configurations, providing insight into the embedding spaces learned by the proposed method and its variants. 
In (a) we depict the embedding for the proposed model. The figure demonstrates a distinct separation between the high cognitive load (orange) and low cognitive load (green) samples. This separation indicates that the model effectively captures the underlying differences in the data, maximizing the inter-class distance while minimizing the intra-class variance. The tight clustering within each class suggests that the model generates highly discriminative representations, leading to robust classification. 
In (b) we present the embedding space of the model without $\mathcal{L}_{OC}$. Here, the UMAP plot shows a noticeable degradation in the separation between classes where the high and low cognitive load samples begin to overlap, particularly along the boundaries. This indicates that $\mathcal{L}_{OC}$ plays a crucial role in enhancing the model’s ability to distinguish between different levels of cognitive load by enforcing a more distinctive embedding space. 
In (c), our model is depicted without the attention mechanism. In this case, the UMAP visualization reveals that while there is still some level of separation between classes, the intra-class compactness is compromised. The high cognitive load samples are more dispersed, and the clusters are less well-defined, indicating the importance of the attention mechanism in generating compact class-specific clusters. 
In (d), we visualize the representations  without both $\mathcal{L}_{OC}$ and attention. In this scenario, the separation between classes is even less distinct, with significant overlap between high and low cognitive load samples. Without these mechanisms, the model struggles to create separate class-specific clusters
leading to less reliable predictions. 
In (e) we remove the multi-spectral topography maps and only rely on the raw data stream, while in (f) we remove the raw data stream and only utilize multi-spectral topography maps. We observe that removal of either domain degrades the quality of the embedding space in comparison to the multi-domain model presented in (a).

\subsection{Sensitivity Analysis}
We conduct sensitivity analyses on a variety of $\beta$ values (see Eq. \ref{total_loss}) with or without the Attention module to determine the most optimal configuration. The $\beta$ hyperparameter adjusts the impact of the $\mathcal{L}_{OC}$ term in the total loss. These configurations and the results are shown in Table \ref{table:sensitivity}. \textcolor{black}{We experimented with multiple $\beta$ values (0.4, 0.7 and 1). $\beta=0.4$ was chosen empirically because it yielded the best average accuracy across all tasks. Higher values (e.g. 0.7 or 1) began to degrade performance.} In Table \ref{table:sensitivity}, for the CL-Drive dataset, when $\beta$ is set to 0.7 with attention, the accuracy decreases by 3.91\%, and the F1 score decreases by 1.23\%. For the same dataset, when $\beta$ is set to 1, the accuracy decreases by 2.70\% and F1 score decreases by 0.82\%. For the CLARE dataset, a $\beta$ value of 0.7 results in a decrease in accuracy by 1.81\% and a decrease in the F1 score by 0.66\%. When $\beta$ is set to 1, the accuracy drops by 4.97\% and the F1 score by 4.8\%.
These results indicate that the use of the Attention module with a $\beta = 0.4$ yields the most effective configuration for the EEG cognitive load classification task. \textcolor{black}{We have conducted a brief ROC curve analysis across different $\beta$ values to demonstrate robustness. As illustrated in Figure S4 in the supplementary file, for a random participant from each dataset, $\beta$ = 0.4 achieves the highest area under the curve (AUC). This demonstrates that this $\beta$ value effectively balances orthogonality enforcement with the model's ability to correctly classify cognitive load.}

\subsection{Robustness Analysis}
To evaluate the robustness of our proposed approach against noise, we conduct an analysis by adding various amounts of Gaussian noise to the test data for both CL-Drive and CLARE datasets. For this experiment, we first calculate the standard deviation of each segment, followed by generating and adding Gaussian noise with 10\%, 30\%, 50\%, and 70\% of the measured standard deviation for each segment. We then use the noisy test samples for evaluating the robustness of our model as well as all the baselines. Figure S1 in the supplementary material
shows the outcome of this experiment on both datasets. 
Specifically, for the CL-Drive dataset (Figure S1 in the supplementary material
(a) and (b)), we can observe that there are considerable drops in accuracy and F1 score when 10\% noise is added for VGG-style, ResNet-style, and MAE networks when trained on features. Meanwhile our method demonstrates slight increase in performance for 10\% noise. Similar trends have been observed in the past with EEG representation learning frameworks where added Gaussian noise has shown to have a positive effect on performance \cite{sampanna2018noise}. This can happen due to a phenomenon known as noise-enhanced signal processing \cite{chapeau2004noise, chen2014noise}, which could occur when weak signals become amplified due to noise, making them easier to distinguish
\cite{wiesenfeld1995stochastic, bulsara1991single, chapeau2004noise}. However, all methods experience performance degradation on this dataset (CL-Drive) with higher amounts of noise, while our method is among the most stable solutions. For CLARE (Figure S1 in the supplementary material
(c) and (d)), we observe a decrease in the performance of our method in terms of accuracy as more and more noise is added while the F1 shows minor degradation. Overall, the results demonstrate that for both datasets, our proposed method shows relatively strong stability compared to most other baseline methods.

\subsection{Band and Channel Importance}
\textcolor{black}{Finally, we analyze the importance of different EEG channels and frequency bands. In this experiment, we zero-mask the values of all channels and bands except for one, and assess the overall performance. Figure S2 in the supplementary material 
(a) and (b) show the individual EEG channel accuracies and F1 scores, (c) and (d) show the individual EEG band accuracies and F1 scores for CL-Drive and CLARE dataset respectively Based on these experiments we notice that all 4 EEG channels are equally important resulting in almost similar accuracy and F1 scores for their respective datasets. Among the 5 EEG frequency bands, we find out that the Theta and Alpha bands give us the highest accuracies and F1 scores for both CL-Drive and CLARE datasets. These observations are well-aligned with prior works such as \cite{liu2023fusion}, where it is been shown that cognitive load is proportional to Theta power, while being inversely proportional to Alpha power. In another prior research \cite{gevins2003neurophysiological}, cognitive load tasks were performed by participants with multiple difficulty levels, where a similar observation was made, indicating an increase in the Theta band power and a decrease in the Alpha band power with high cognitive load. This suggests that both of these bands are strong indicators of cognitive load. For the CLARE dataset, we notice an increase in accuracy for the Beta frequency band, which could be due to voluntary hand movements while performing the task. In \cite{munoz2013analysis}, it was demonstrated that Alpha and Beta frequency bands show increased activity during both actual and imagined movements.}

\subsection{\textcolor{black}{Attention Weight Distribution}}
\textcolor{black}{We extracted attention weights from the trained fusion layer across all test samples to visualize the modality selection pattern. In Figure S3 in the supplementary file, we show the heatmap illustration of the learned modality preferences across the embedding dimensions for 10 random samples. Our analysis reveals a dimension-specific modality selection, with attention weights leaning toward either topographical maps shown in yellow, or raw EEG data shown in purple, rather than a uniform distribution. This indicates that the model has successfully learned to identify which modality provides the most discriminative information for each feature dimension, hence optimizing the fusion process. This further validates our approach over fixed fusion strategies.}

\subsection{\textcolor{black}{Real-time Deployment Feasibility}}
\textcolor{black}{To address the real-time deployment feasibility, we conducted comprehensive inference time analysis across all participants (Table S2 in the supplementary file). The results demonstrate that our model achieves a mean total inference time of 20.92 ± 8.20 ms across all participants, substantially below the 100 ms threshold required for safety-critical applications such as real-time driver cognitive state monitoring. Component-wise analysis reveals that the encoders dominate the computation time, with the $Enc_\text{topo}$ requiring 6.11 ± 7.18 ms and the $Enc_\text{raw}$ requiring 13.31 ± 4.48 ms. The fusion mechanism introduces minimal overhead (1.47 ± 0.37 ms), indicating that our attention-based adaptive weighting strategy does not compromise computational efficiency. }

\section{Conclusion and Future Work} \label{Conclusion&FutureWork}
We proposed a novel multi-domain approach to classify cognitive load from EEG. Our method leverages representations in both time and frequency domains using a dual-stream network. The utilization of a multi-domain attention module allows for effective fusion of relevant features across the two domains, thus enhancing the representations that are critical for accurate cognitive load classification. Furthermore, the implementation of an orthogonal loss aids in enhancing distinctions between classes. Our method demonstrates notable improvements in performance over state-of-the-art solutions as well as single-domain variants, on two public datasets, CL-Drive and CLARE. Detailed ablation, sensitivity, and robustness studies demonstrate the important effect of each main component of our method, as well as stability under noisy inputs. We finally evaluated the importance of individual EEG channels and frequency bands, showing that Theta and Alpha bands have the highest influence on cognitive load compared to others while all four channels result in similar performances.

\textcolor{black}{Given that our approach uses both time and frequency domain information, it is more complex than single-domain models. \textcolor{black}{Our fusion operates at the feature level and does not provide fine-grained spatial (channel-wise) or temporal attention. Implementing such mechanisms would significantly increase computational complexity and potentially compromise real-time performance. Future work could explore lightweight spatial-temporal attention that balances these.} Moreover, the weight for the orthogonality constraint ($\beta$) requires tuning to balance representational diversity against convergence stability. Our ablation studies (Table \ref{table:sensitivity}) indicate that inappropriate $\beta$ values can degrade performance. Finally, while our results show clear improvements on the studied dataset, the generalizability of the approach to very different EEG tasks remains to be explored in future work.}


\end{bibunit}



\clearpage
\onecolumn
\appendix
\section*{Supplementary Material}
\vspace{5mm}
\begin{bibunit}[IEEEtran]







\begin{figure}[h!]
    \centering
    \begin{subfigure}[t]{0.23\linewidth}
        \includegraphics[width=\linewidth]{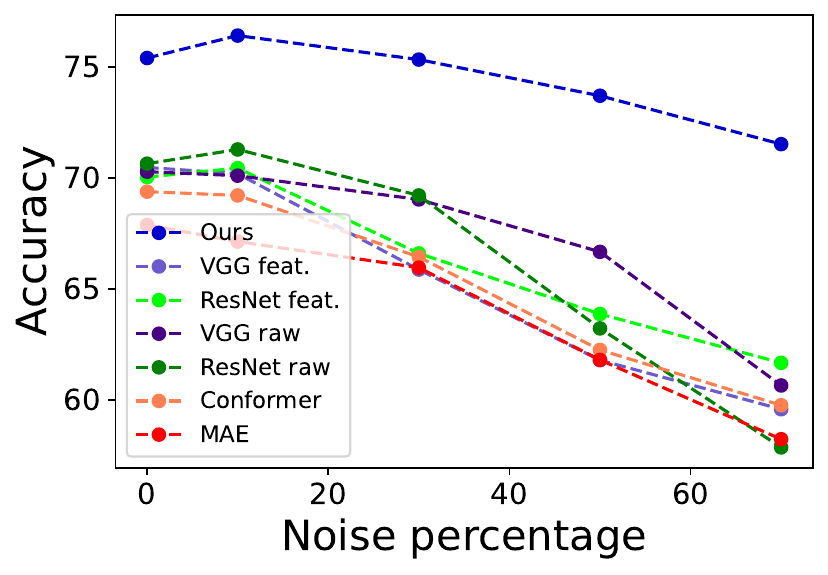}
        \caption{\small Acc. vs. noise, CL-Drive}
    \end{subfigure}
    \hfill
    \begin{subfigure}[t]{0.23\linewidth}
        \includegraphics[width=\linewidth]{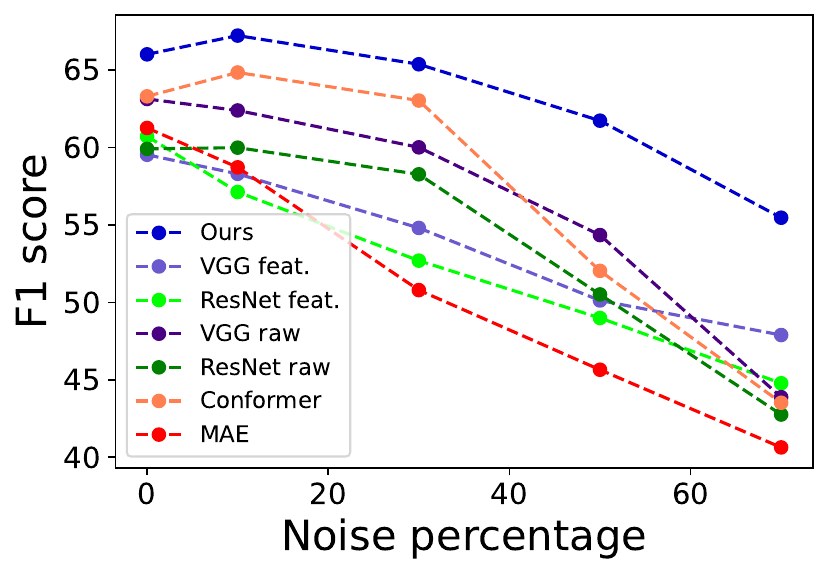}
        \caption{\small F1 vs. noise, CL-Drive}
    \end{subfigure}
    \hfill
    \begin{subfigure}[t]{0.23\linewidth}
        \includegraphics[width=\linewidth]{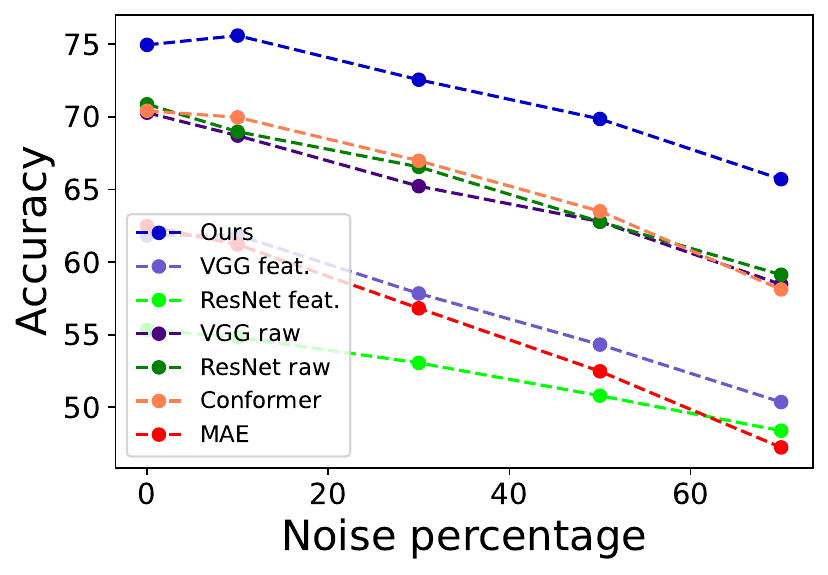}
        \caption{\small Acc. vs. noise, CLARE}
    \end{subfigure}
    \hfill
    \begin{subfigure}[t]{0.23\linewidth}
        \includegraphics[width=\linewidth]{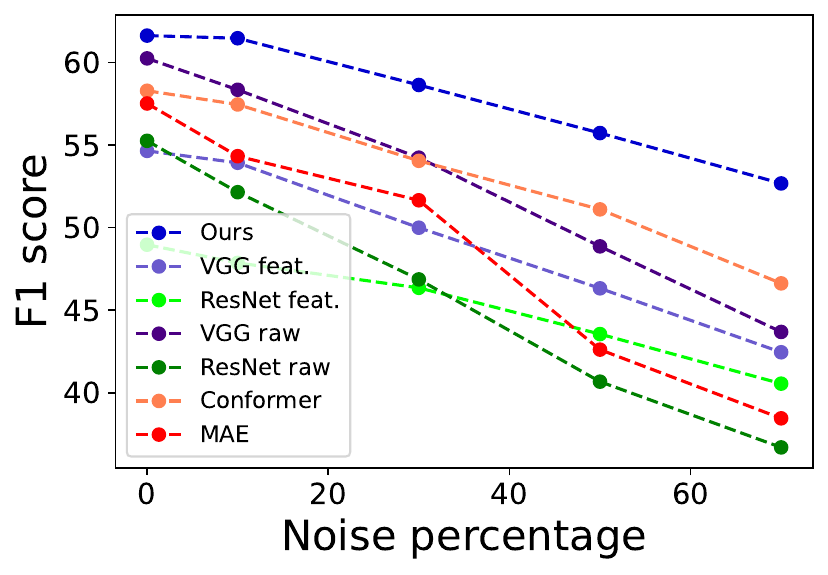}
        \caption{\small F1 vs. noise, CLARE}
    \end{subfigure}
    \caption{Accuracy and F1 scores for our method vs. baselines with increasing Gaussian noise.}
    \label{fig:gaussian_noise}
\end{figure}

\vspace{1em}

\begin{figure}[h!]
    \centering
    \begin{subfigure}[t]{0.23\linewidth}
        \includegraphics[width=\linewidth]{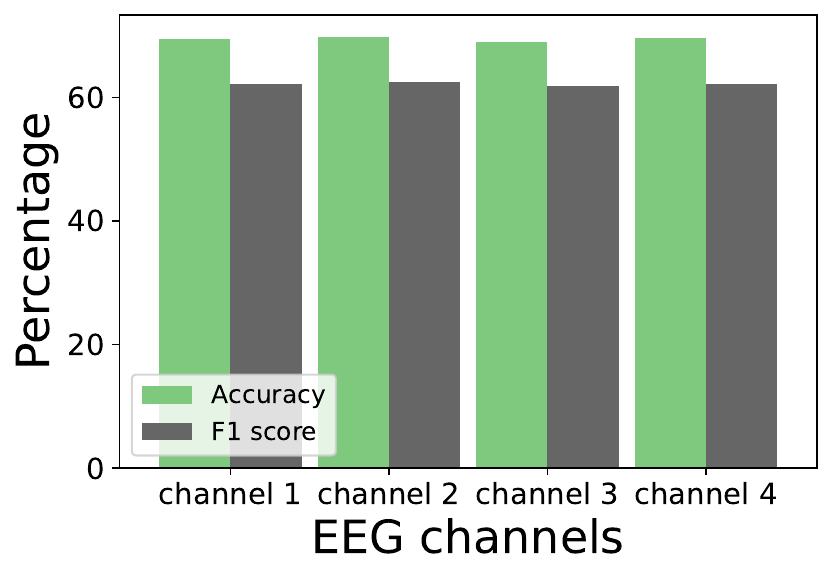}
        \caption{\small Channels, CL-Drive}
    \end{subfigure}
    \hfill
    \begin{subfigure}[t]{0.23\linewidth}
        \includegraphics[width=\linewidth]{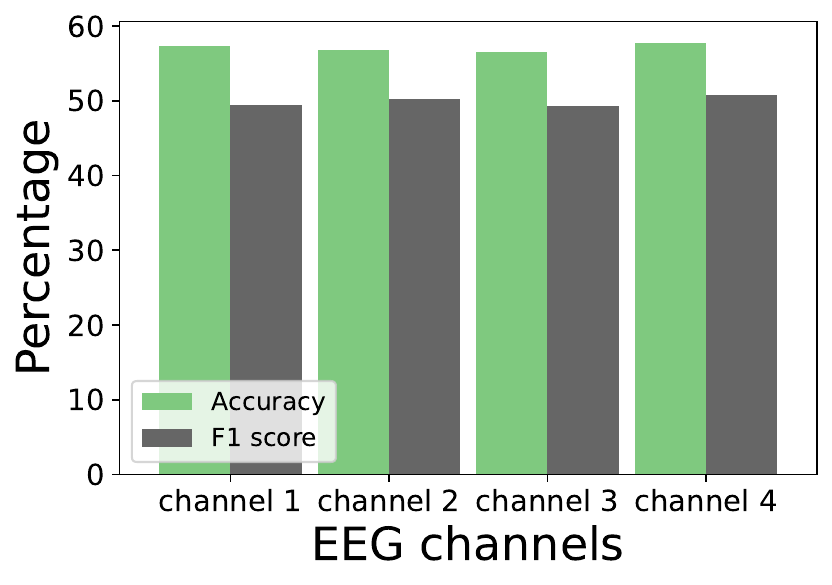}
        \caption{\small Channels, CLARE}
    \end{subfigure}
    \hfill
    \begin{subfigure}[t]{0.23\linewidth}
        \includegraphics[width=\linewidth]{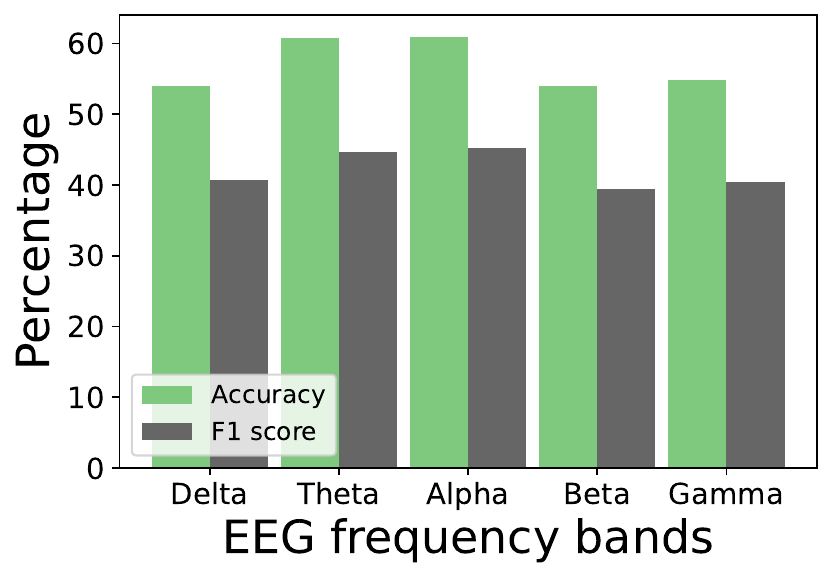}
        \caption{\small Freq. bands, CL-Drive}
    \end{subfigure}
    \hfill
    \begin{subfigure}[t]{0.23\linewidth}
        \includegraphics[width=\linewidth]{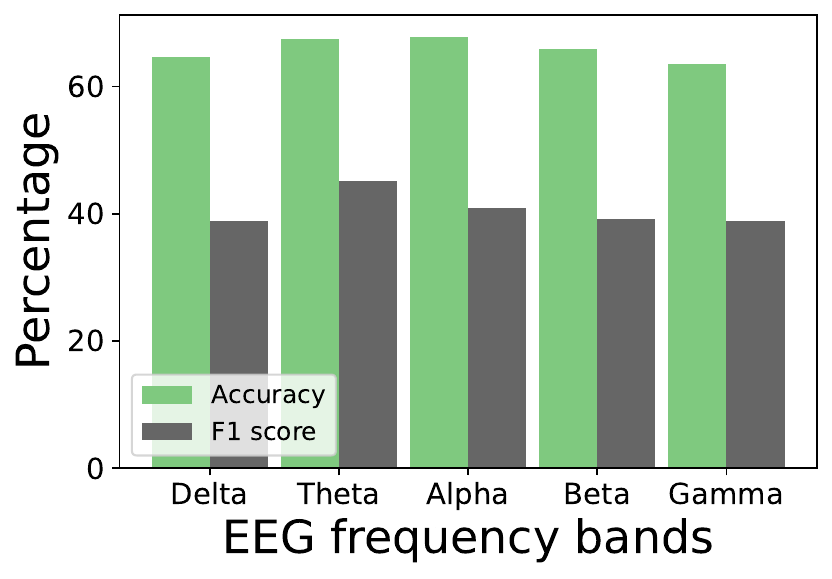}
        \caption{\small Freq. bands, CLARE}
    \end{subfigure}
    \caption{Accuracy and F1 scores across EEG channels and frequency bands.}
    \label{fig:band_ch_importance}
\end{figure}


\vspace{1em}

\begin{figure}[h!]
    \centering
    \begin{subfigure}[t]{0.48\linewidth}
        \includegraphics[width=\linewidth]{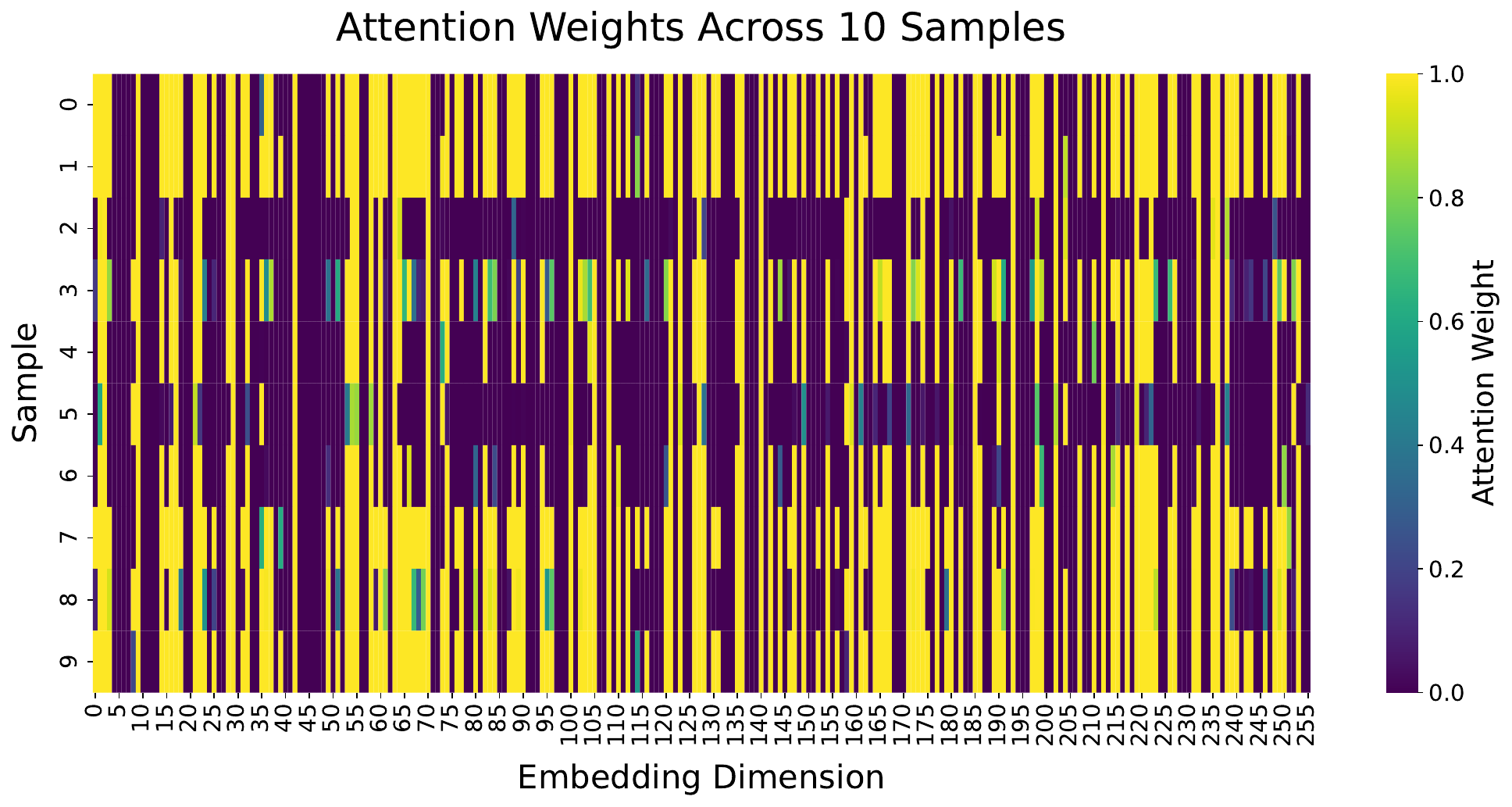}
        \caption{\small CL-Drive}
    \end{subfigure}
    \hfill
    \begin{subfigure}[t]{0.48\linewidth}
        \includegraphics[width=\linewidth]{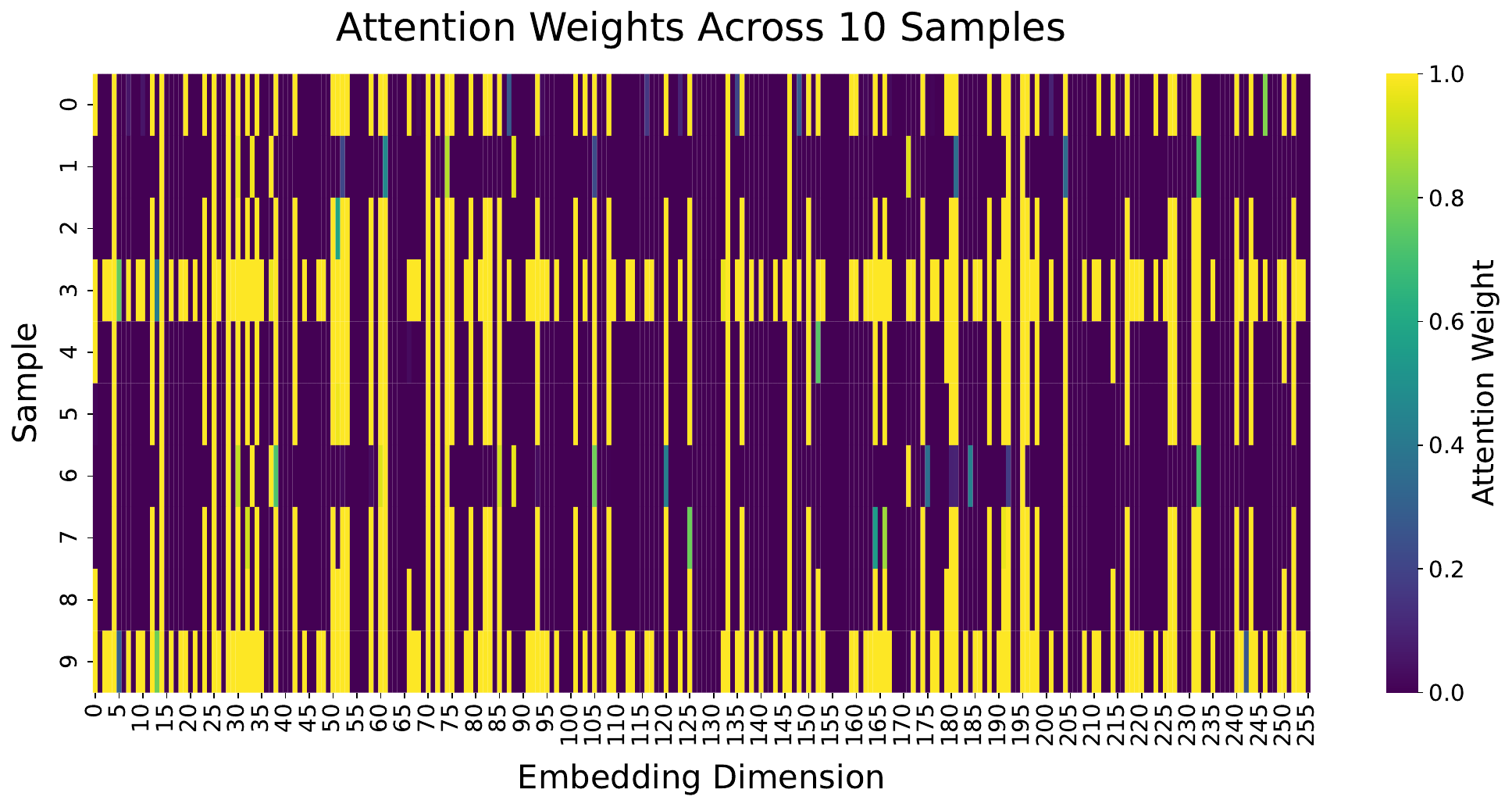}
        \caption{\small CLARE}
    \end{subfigure}
    \hfill
    \caption{Attention weights of the fusion layer across embedding dimensions for 10 samples from (a) CL-Drive and (b) CLARE datasets. Yellow indicates topographic image dominance, purple indicates raw EEG dominance.}
    \label{fig:attention_heatmaps}
\end{figure}

\clearpage

\vspace{1em}


\begin{figure}[h!]
    \centering
    \begin{subfigure}[t]{0.48\linewidth}
        \includegraphics[width=\linewidth]{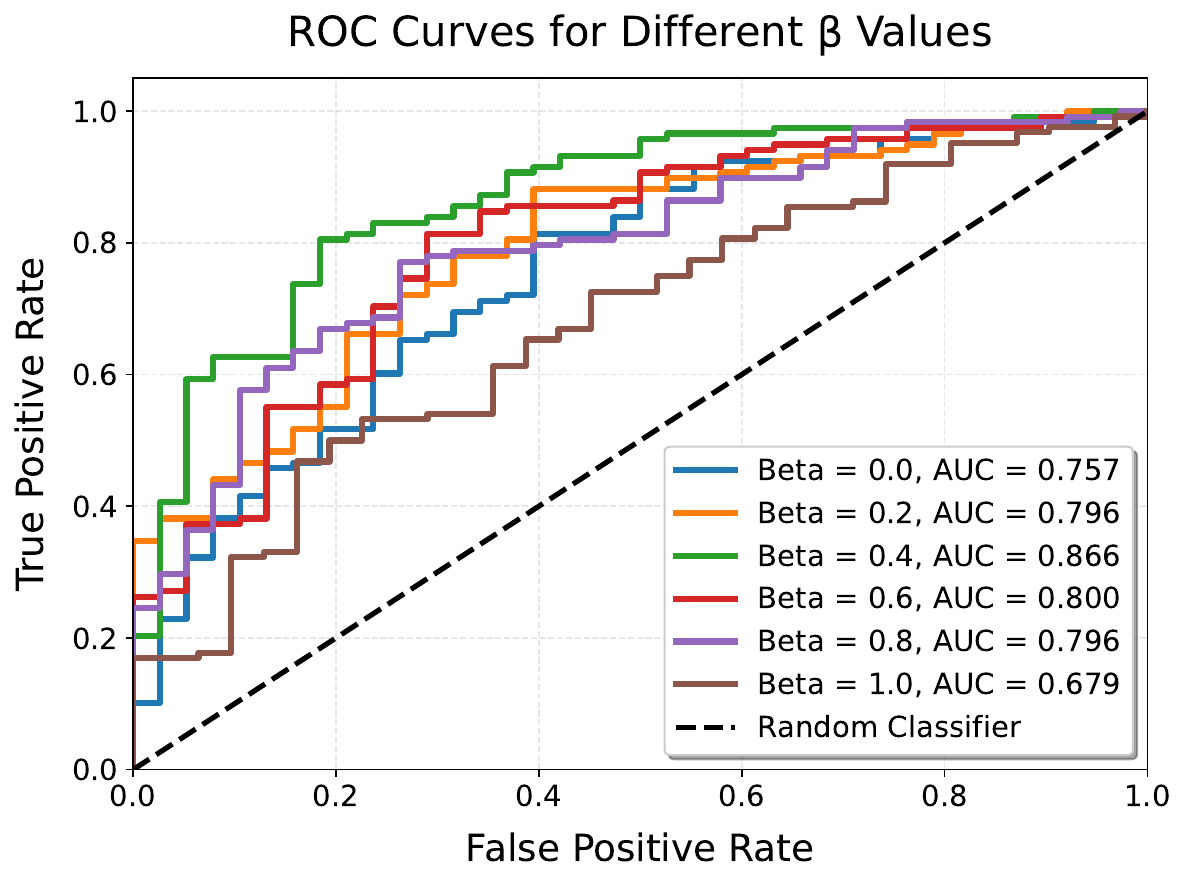}
        \caption{\small CL-Drive}
    \end{subfigure}
    \hfill
    \begin{subfigure}[t]{0.48\linewidth}
        \includegraphics[width=\linewidth]{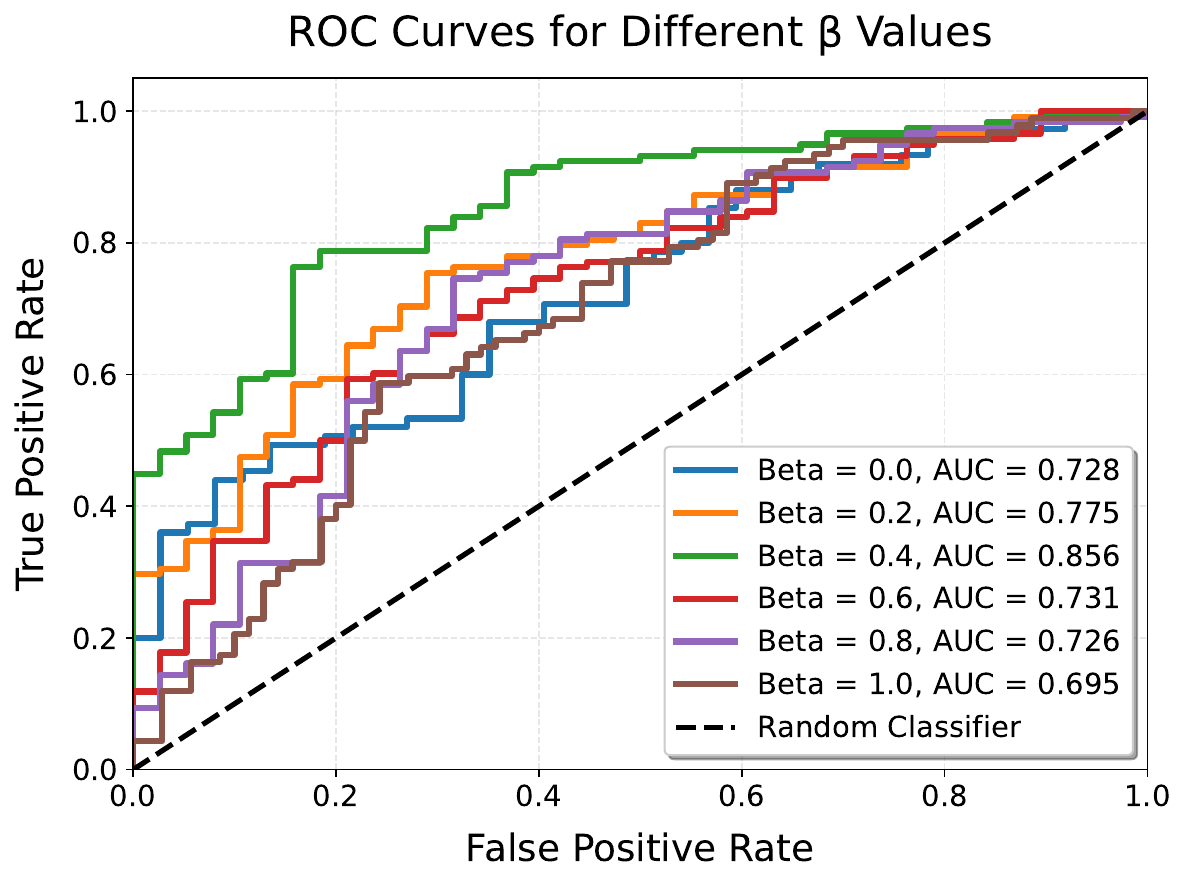}
        \caption{\small CLARE}
    \end{subfigure}
    \hfill
    \caption{ROC curves demonstrating the impact of $\beta$ on classification performance for a random participant from (a) CL-Drive and (b) CLARE datasets. $\beta$ = 0.4 achieves the highest AUC values (0.866 for CL-Drive, 0.853 for CLARE)}
    \label{fig:beta}
\end{figure}

\vspace{0.5em}

\noindent
\textbf{Table S1:} {Performance of our proposed method compared to several state-of-the-art solutions using multi-spectral topography maps using DE features instead of PSD.}

\vspace{0.5em}

\small
\setlength{\tabcolsep}{8pt}
\begin{center}
\begin{tabular}{l|ll|ll}
     \hline
    \multicolumn{1}{c|}{} &
    \multicolumn{2}{c|}{CL-Drive} & \multicolumn{2}{c}{CLARE} \\
    Model  &  Accuracy & F1 score &  Accuracy & F1 score \\
    \hline\hline
    MAE \cite{pulver2023eeg}  & 
    63.78(18.98) & 57.21(17.49) & 58.17(12.67) & 53.67(11.51) \\
    ResNet-style \cite{angkan2024multimodal} &
    66.26(20.37) & 56.10(16.48) & 56.78(17.34) & 46.18(11.88) \\
    VGG-style \cite{angkan2024multimodal} &
    67.16(18.53) & 55.93(11.31) & 59.18(15.24) & 51.00(10.28) \\
    MMCosine \cite{xu2023mmcosine} &  
    72.74(9.05) & 63.40(11.21) & 69.33(12.68) & 58.49(10.89) \\
    Proposed  & 
    \textbf{73.09(8.08)} & \textbf{64.99(8.89)} & 
    \textbf{72.12(14.53)} & \textbf{58.86(12.42)} \\
    \hline
\end{tabular}
\end{center}

\vspace{0.5em}


\noindent
\textbf{Table S2:} {Component-wise inference time analysis (ms) showing mean and standard deviation for each model component across all participants.}

\vspace{0.5em}

\small
\setlength{\tabcolsep}{8pt}
\centering
\begin{tabular}{l|l|l|l|l}
     \hline
    & Mean(std) & Min(std) & Max(std) & Median(std) \\
    \hline\hline 
    $Enc_\text{topo}$ & 6.11(7.18) & 4.91(5.28)	& 7.25(8.52) & 6.13(7.64) \\
    $Enc_\text{raw}$ & 13.31(4.48) &	11.97(4.23) & 15.02(4.95) &	13.04(4.46) \\
    Fusion & 1.47(0.37) & 1.09(0.29) & 1.95(0.78) &	1.45(0.46) \\
    Total & 20.92(8.20) & 19.05(7.31) & 22.99(8.87)	& 20.73(8.70) \\
    
    \hline
\end{tabular}




\vspace{5mm}
\begin{center}

\end{center}
\end{bibunit}



\vfill

\end{document}